\pdfoutput=1
\documentclass[11pt,twoside,a4paper,cmspaper,final,collab]{cms-tdr}
\def\svnVersion{e4a0950}\def\svnDate{2024/07/31}\def\cmsCernNoTag{CERN-EP-2024-049}\def\cmsCernDate{\today}\def\cmsMessage{Published in the Journal of High Energy Physics as \href{http://dx.doi.org/10.1007/JHEP07(2024)142}{\doi{10.1007/JHEP07(2024)142}.}}
\begin{document}\cmsNoteHeader{EXO-22-015}

\newlength\cmsTabSkip\setlength{\cmsTabSkip}{1ex}

\newcommand{\cmax}{\ensuremath{^\text{max}}\xspace}
\newcommand{\cmin}{\ensuremath{^\text{min}}\xspace}
\newcommand{\DN}{\ensuremath{D_{N}}\xspace}
\newcommand{\DNmax}{\ensuremath{\DN\cmax}\xspace}
\newcommand{\Rcut}{\ensuremath{R_\text{max}}\xspace}
\newcommand{\trkdxy}{\ensuremath{d_{xy}}\xspace}
\newcommand{\trkdxycmin}{\ensuremath{\trkdxy\cmin}\xspace}
\newcommand{\ipxy}{\trkdxy}
\newcommand{\dxysig}{\ensuremath{\frac{\ipxy}{\sigma(\ipxy)}}\xspace}
\newcommand{\trkdz}{\ensuremath{d_{z}}\xspace}
\newcommand{\ipz}{\ensuremath{d_{z}}\xspace}
\newcommand{\ipzmax}{\ensuremath{\ipz\cmax}\xspace}
\newcommand{\taun}[1]{\ensuremath{\overline{\tau}_{#1}}\xspace}
\newcommand{\taunn}{\ensuremath{\taun{n}}\xspace}
\newcommand{\med}[1]{\ensuremath{\left\langle #1 \right\rangle}}
\newcommand{\medipxy}{\ensuremath{\med{\ipxy}}\xspace}
\newcommand{\alphd}{\ensuremath{\alpha_\text{3D}}\xspace}
\newcommand{\Nc}{\ensuremath{N_{\text{c}}}\xspace}
\newcommand{\dark}{\text{dark}}
\newcommand{\Ncdark}{\ensuremath{N_{\text{color}}^\dark}\xspace}
\newcommand{\Nfdark}{\ensuremath{N_{\text{flavor}}^\dark}\xspace}
\newcommand{\Lamdark}{\ensuremath{\Lambda_\dark}\xspace}
\newcommand{\dquark}{\ensuremath{\text{Q}_\dark}\xspace}
\newcommand{\dpi}{\ensuremath{\pi_\dark}\xspace}
\newcommand{\dmed}{\ensuremath{\text{X}_\dark}\xspace}
\newcommand{\dcoupl}{\ensuremath{\kappa_{\alpha i}}\xspace}
\newcommand{\ctau}{\ensuremath{c\tau}\xspace}
\newcommand{\ctaudpi}{\ensuremath{\ctau_{\dpi}}\xspace}
\newcommand{\ctaumax}{\ensuremath{\ctaudpi^{\text{max}}}\xspace}
\newcommand{\kappaZero}{\ensuremath{\kappa_0}\xspace}
\newcommand{\sigmaopt}{\ensuremath{\sigma_\text{opt}}\xspace}

\newcommand{\enc}[1]{\ensuremath{\left( #1 \right)}}

\newcommand{\mi}{\ensuremath{m_{i}}\xspace}
\newcommand{\mj}{\ensuremath{m_{j}}\xspace}
\newcommand{\mdown}{\ensuremath{m_{\PQd}}\xspace}
\newcommand{\mdpi}{\ensuremath{m_{\dpi}}\xspace}
\newcommand{\fdpi}{\ensuremath{f_{\dpi}}\xspace}
\newcommand{\mdmed}{\ensuremath{m_{\dmed}}\xspace}

\newcommand{\abseta}{\ensuremath{\abs{\eta}}\xspace}
\newcommand{\gjets}{\ensuremath{\PGg{+}\text{jets}}\xspace}
\newcommand{\deta}{\ensuremath{\Delta\eta}\xspace}
\newcommand{\dphi}{\ensuremath{\Delta\phi}\xspace}
\newcommand{\dR}{\ensuremath{\Delta R}\xspace}
\newcommand{\ptsup}[1]{\ensuremath{p_{\mathrm{T}}^{#1}}\xspace}
\newcommand{\pttrack}{\ensuremath{\ptsup{\text{track}}}\xspace}
\newcommand{\ntrack}{\ensuremath{n_\text{track}}\xspace}
\newcommand{\Fb}[1]{\ensuremath{F_\PQb^{\text{#1}}}\xspace}
\newcommand{\muF}{\ensuremath{\mu_\text{F}}\xspace}
\newcommand{\muR}{\ensuremath{\mu_\text{R}}\xspace}
\newcommand{\DEEPJET}{\textsc{DeepJet}\xspace}
\newcommand{\epsi}{\ensuremath{\epsilon_{i}}\xspace}
\newcommand{\epsj}{\ensuremath{\epsilon_{j}}\xspace}
\newcommand{\epsk}{\ensuremath{\epsilon_{k}}\xspace}
\newcommand{\epsavg}{\ensuremath{\epsilon^{\text{avg}}}\xspace}
\newcommand{\epsf}{\ensuremath{\epsilon(f, \pt)}\xspace}
\newcommand{\epsb}{\ensuremath{\epsilon(\PQb, \pt)}\xspace}
\newcommand{\epsq}{\ensuremath{\epsilon(\PQq, \pt)}\xspace}

\newenvironment{cmsequation}{\begin{linenomath}\begin{equation}}{\end{equation}\end{linenomath}\ignorespacesafterend}
\newenvironment{cmsequation*}{\begin{linenomath}\begin{equation*}}{\end{equation*}\end{linenomath}\ignorespacesafterend}

\newcommand{\llapscript}[1]{\llap{$\scriptstyle#1$}}
\newcommand{\leftscript}[2]{\ensuremath{\phantom{{}_{#1}^{#2}}{\vphantom{0}}_{\llapscript{#1}}^{\llapscript{#2}}}}
\newcommand{\asymdelim}{$\,\vphantom{0}_{-}^{+}\,$}
\newcommand{\symmdelim}{$\,\pm\,$}
\newcommand{\TabHeaderA}{Estimation}
\newlength{\TabLengthA}
\settowidth{\TabLengthA}{\TabHeaderA}
\newcommand{\TabHeaderB}{stat.}
\newlength{\TabLengthB}
\settowidth{\TabLengthB}{\TabHeaderB}
\newcommand{\TabHeaderC}{syst.}
\newlength{\TabLengthC}
\settowidth{\TabLengthC}{210.210}
\newcolumntype{A}{>{\raggedleft\arraybackslash}m{\dimexpr.85\TabLengthA}@{\hspace{0em}}m{\dimexpr.2\TabLengthA}}
\newcolumntype{B}{>{\raggedleft\arraybackslash}m{\dimexpr.5\TabLengthB}@{\hspace{0em}}m{\dimexpr.5\TabLengthB}}
\newcolumntype{C}{>{\raggedleft\arraybackslash}m{\dimexpr.5\TabLengthC}@{\hspace{0em}}m{\dimexpr.5\TabLengthC}}
\newcolumntype{Z}{A@{\asymdelim}B@{\symmdelim}C}
\newlength{\TabLengthG}
\settowidth{\TabLengthG}{\TabHeaderC}
\newlength{\TabLengthF}
\settowidth{\TabLengthF}{\TabHeaderB.}
\newcolumntype{E}{>{\raggedleft\arraybackslash}m{\dimexpr.75\TabLengthA}@{\hspace{0em}}m{\dimexpr.3\TabLengthA}}
\newcolumntype{F}{>{\raggedleft\arraybackslash}m{\dimexpr.45\TabLengthF}@{\hspace{0em}}m{\dimexpr.55\TabLengthF}}
\newcolumntype{G}{>{\raggedleft\arraybackslash}m{\dimexpr.5\TabLengthG}@{\hspace{0em}}m{\dimexpr.5\TabLengthG}}
\newcolumntype{H}{E@{\asymdelim}F@{\symmdelim}G}
\newcommand{\TabHeaderO}{Observed yield}
\newlength{\TabLengthO}
\settowidth{\TabLengthO}{\TabHeaderO}
\newcolumntype{O}{>{\raggedleft\arraybackslash}m{\TabLengthO}<{\hphantom{.yield}\arraybackslash}}
\newcolumntype{Q}{>{\raggedleft\arraybackslash}m{\TabLengthO}<{\hphantom{. yield}\arraybackslash}}
\newcommand{\TabHeadSpace}{\rule{0pt}{2.6ex}}

\cmsNoteHeader{EXO-22-015}
\title{Search for dark QCD with emerging jets in proton-proton collisions at \texorpdfstring{$\sqrt{s} = 13\TeV$}{sqrt(s) = 13 TeV}}

\author*[cern]{The CMS Collaboration}

\date{\today}

\abstract{
   A search for ``emerging jets'' produced in proton-proton collisions at a center-of-mass energy of 13\TeV is performed using data collected by the CMS experiment 
   corresponding to an integrated luminosity of 138\fbinv. This search examines a hypothetical dark quantum chromodynamics (QCD) sector that couples to the standard model (SM) through a scalar mediator. The scalar mediator decays into an SM quark and a dark sector quark. As the dark sector quark showers and hadronizes, it produces long-lived dark mesons that subsequently decay into SM particles, resulting in a jet, known as an emerging jet, with multiple displaced vertices. This search looks for pair production of the scalar mediator at the LHC, which yields events with two SM jets and two emerging jets at leading order. The results are interpreted using two dark sector models with different flavor structures, and exclude mediator masses up to 1950 (1950)\GeV for an unflavored (flavor-aligned) dark QCD model. 
   The unflavored results surpass a previous search for emerging jets by setting the most stringent mediator mass exclusion limits to date, while the flavor-aligned results provide the first direct mediator mass exclusion limits to date.
}

\hypersetup{%
pdfauthor={CMS Collaboration},%
pdftitle={Search for dark QCD with emerging jets in proton-proton collisions at sqrt(s) = 13 TeV},%
pdfsubject={CMS},%
pdfkeywords={CMS, emerging jets, dark QCD, dark matter}}

\maketitle 

\section{Introduction}
\label{sec:intro}

Although there is a preponderance of evidence  from astronomical and cosmological observations~\cite{Rubin:1980zd,Persic:1995ru,Clowe:2006eq,Chang:2017kmv,Planck:2018vyg} for the existence of dark matter (DM), it has not yet been detected in laboratories, suggesting that its origin may be associated with as-of-yet unobserved physics processes beyond the standard model (SM). 
As experimental searches have excluded a large portion of the phase space of DM models with weakly interacting massive particles, alternative theoretical models have been developed with a hidden gauge sector, some of which are similar to quantum chromodynamics (QCD), which can result in strongly self-interacting DM particles~\cite{Strassler:2006im,Petraki:2013wwa,Beauchesne:2018myj,Bai:2013xga}. Dark matter of this type could interact with SM particles through so-called mediator particles and potentially be produced at colliders, generating signatures such as semivisible jets~\cite{Cohen:2015toa} or emerging jets (EJs)~\cite{Schwaller:2015gea}. 

\begin{figure}[htbp]
    \centering
    \includegraphics[width=0.49\textwidth]{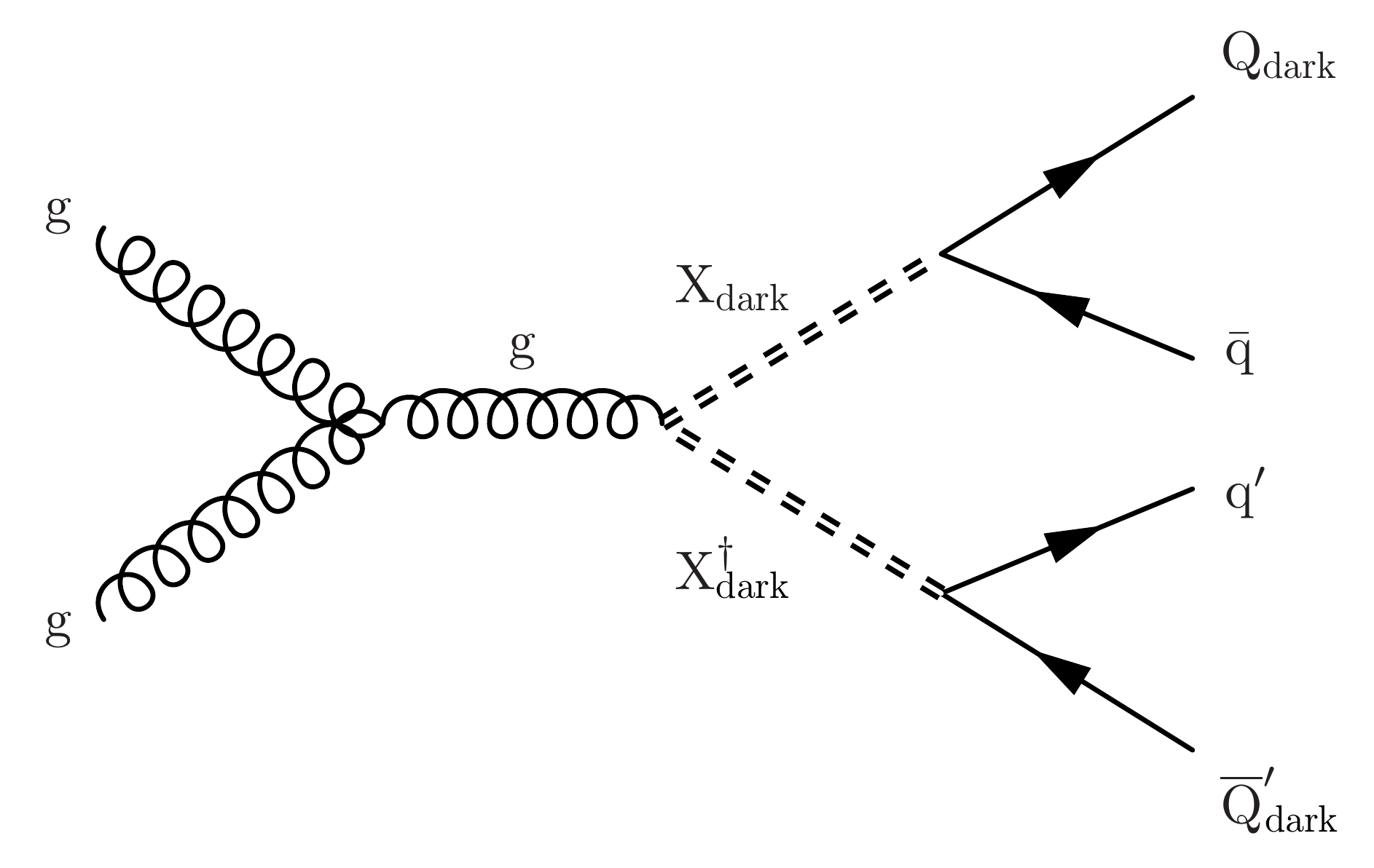}
    \includegraphics[width=0.49\textwidth]{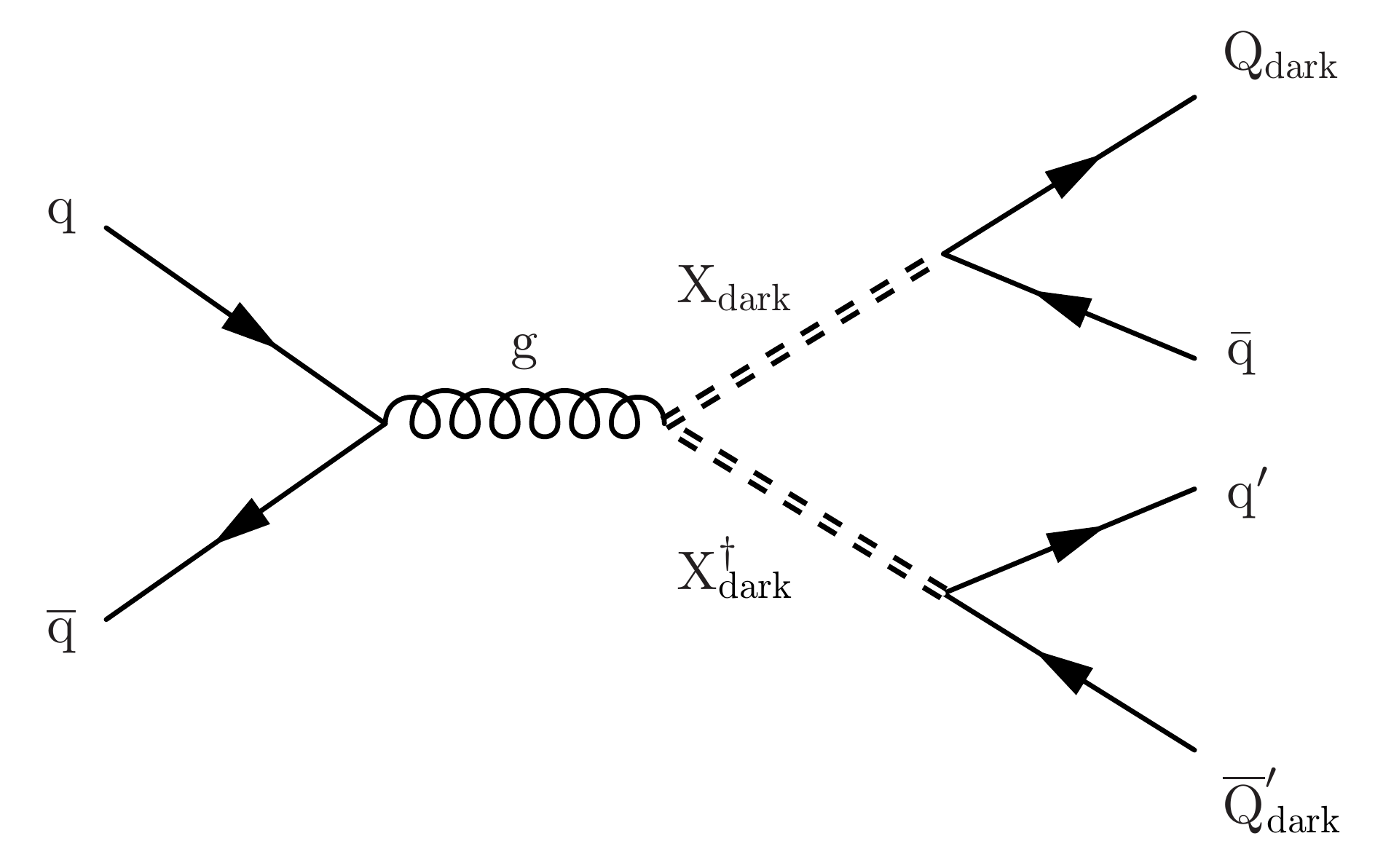} 
    \caption{Feynman diagrams for pair production of dark mediator particles via gluon-gluon fusion (left) and quark-antiquark annihilation (right), with each mediator decaying to an SM quark and a dark quark.}
    \label{fig:diagram_emj}
\end{figure}

The search described in this paper is motivated by the models proposed in Refs.~\cite{Schwaller:2015gea,Agrawal:2014aoa,Renner:2018fhh}. 
A composite dark sector with a QCD-like non-Abelian gauge symmetry $\text{SU}(\Ncdark)$, where \Ncdark is the number of dark colors, is added to the SM gauge group. We consider the case where fermions in the dark sector (dark quarks \dquark) interact with the SM quarks through a scalar mediator \dmed. The scalar mediator is charged under both SM QCD and dark QCD and couples to a quark and a dark quark via Yukawa interactions
with coupling strength \dcoupl, where the subscript $\alpha$ ($i$) denotes flavors of dark (SM) quarks. The mediator can be pair produced through gluon splitting at the LHC, similar to the pair production of a single type of squark  in supersymmetry~\cite{Fayet:1976cr}. In fact, the production cross section is the same as for  pair production of right-handed top squarks~\cite{Borschensky:2014cia,SUSYProd:Beenakker:2016lwe} multiplied by \Ncdark. Each mediator decays into a quark and a dark quark, as shown in Fig.~\ref{fig:diagram_emj}. The dark quarks hadronize into dark hadrons. Generally, of the dark hadrons in the composite QCD-like dark sector described above, either the lightest dark baryon or dark meson is stable and thus provides a natural DM candidate.

The first version of this model that we consider, referred to as ``unflavored'', has a simplified flavor structure in the coupling of the dark sector to SM particles, such that only the Yukawa coupling to the d quark is nonnegligible~\cite{Schwaller:2015gea}.
Equation~\eqref{eq:ctau_unflavored}, taken from Ref.~\cite{Schwaller:2015gea}, gives the proper decay length of dark pions in this model:
\begin{cmsequation}
    \ctaudpi =
    80\mm 
    \enc{\frac{1}{\kappa^4}} 
    \enc{\frac{2\GeV}{\fdpi}}^2
    \enc{\frac{100\MeV}{\mdown}}^2
    \enc{\frac{2\GeV}{\mdpi}}
    \enc{\frac{\mdmed}{1\TeV}}^4,
    \label{eq:ctau_unflavored}
\end{cmsequation}
where $\kappa$ is the coupling between dark quarks and the SM down quark, \fdpi is the dark pion decay constant, \mdown is the mass of the SM down quark, and \mdpi is the dark pion mass. The numbers with units show the expected typical scale for each variable.

If instead multiple Yukawa couplings \dcoupl have nonnegligible values, the proper decay length for dark mesons composed of dark quarks of flavor indices $\alpha$ and $\beta$ is given by Eq.~\eqref{eq:ctau_flavored} from Ref.~\cite{Renner:2018fhh}:
\begin{cmsequation}
    \ctau_{\dpi}^{\alpha\beta} = \frac{8\pi \mdmed^{4}c\hbar}{
    \Nc \mdpi \fdpi^2 \displaystyle{\sum\limits_{i,j}}
    \abs{\dcoupl\kappa_{\beta j}^*}^2 
    \enc{\mi^2+\mj^2}
    \sqrt{\!
        \enc{1-\frac{(\mi+\mj)^2}{\mdpi^2}}\!
        \enc{1-\frac{(\mi-\mj)^2}{\mdpi^2}}
    }},
    \label{eq:ctau_flavored}
\end{cmsequation}
where \mdmed is the mediator particle mass, \Nc is the SM color factor, and \mi, \mj are the masses of the SM quarks with flavor indices $i$, $j$, respectively. Within this model, we focus on the ``flavor-aligned'' scenario from Ref.~\cite{Renner:2018fhh}, which has three dark quark flavors that couple to the SM down-type quarks (\PQd, \PQs, \PQb) via a diagonal matrix $\dcoupl = \kappaZero \delta_{\alpha i}$. Because of spin considerations, dark hadron decays to heavier SM particles are favored, typically resulting in a large number of \PQb quarks in the decays when kinematically allowed. We characterize the flavor-aligned model in terms of the maximum proper lifetime for any dark pion, \ctaumax.
Following the benchmark models from Refs.~\cite{Schwaller:2015gea, Renner:2018fhh}, where the lightest dark baryon is stable, and given that dark baryon production is suppressed by $1/\Ncdark$~\cite{Witten:1979kh}, we consider only dark meson production and require all dark mesons to decay finally into SM particles.
When the lifetimes are long enough to give macroscopic decay distances,  
the resulting signature from either model is two SM jets and two EJs containing multiple displaced vertices.

Previous experimental searches for strongly self-interacting DM have been made using proton-proton (pp) collision data collected at the CERN LHC at $\sqrt{s}=13\TeV$ and corresponding to about 140\fbinv.  This includes a search for semivisible jets by the CMS Collaboration~\cite{CMS:2021dzg}, a search for semivisible jets by the ATLAS Collaboration~\cite{SVJ:ATLAS:2023swa}, and a search for dark quarks in a dijet final state by ATLAS~\cite{ATLAS:2023kao}.
In addition, a previous search for EJs interpreted with an unflavored dark sector model was performed by the CMS Collaboration using a data sample collected in 2016, corresponding to an integrated luminosity of 16.1\fbinv~\cite{CMSEMJ:Sirunyan:2018njd}. 

\begin{sloppypar}In this paper, we present a search for the EJ signatures of the unflavored and the flavor-aligned dark sector models, focusing on the displacement information of the tracks in the EJ, using the data set collected by the CMS Collaboration in 2016--2018 using pp collisions at a center-of-mass energy of 13\TeV and corresponding to 138\fbinv. The search strategy first identifies EJs by exploiting their topological differences relative to SM jets using selection criteria optimized separately for each model. Then, the probability for an SM jet to be misidentified as an EJ is measured, and the background is estimated using control samples in data. The main background comes from SM multijet production where SM jets are misidentified as EJs.\end{sloppypar}

Compared to the previous analysis~\cite{CMSEMJ:Sirunyan:2018njd}, we have extended the unflavored EJ model search to a wider parameter space. This study also includes the first dedicated search for a flavored dark QCD sector. We have implemented a number of important changes that considerably increase the sensitivity of the search, including the incorporation of a machine learning (ML) technique aimed at enhancing the EJ identification (tagging) performance. The tabulated results are provided as a HEPData record~\cite{HEPData}.

This paper is organized as follows. In Section~\ref{sec:cms} we give an introduction of the CMS detector, followed in Section~\ref{sec:mc} by a detailed description of the simulated data used in this search. The event reconstruction and triggering algorithms are discussed in Section~\ref{sec:reco}. In Section~\ref{sec:method} we present the analysis strategy and two independent EJ tagging methods. Section~\ref{sec:bkg} describes the background estimation method. The treatment of uncertainties is detailed in Section~\ref{sec:unc}.  The results are presented in Section~\ref{sec:results} and summarized in Section~\ref{sec:conclusion}.

\section{The CMS detector}
\label{sec:cms}

The central feature of the CMS apparatus is a superconducting solenoid of 6\unit{m} internal diameter, providing a magnetic field of 3.8\unit{T}. Located within the solenoid volume are a silicon pixel and strip tracker, a lead tungstate crystal electromagnetic calorimeter (ECAL), and a brass and scintillator hadron calorimeter (HCAL), each composed of a barrel and two endcap sections. Forward calorimeters extend the pseudorapidity $(\eta)$ coverage provided by the barrel and endcap detectors. Muons are measured in gas-ionization detectors embedded in the steel flux-return yoke outside the solenoid. 

The silicon tracker used in 2016 measured charged particles within the range $\abseta < 2.5$. For nonisolated particles of $1 < \pt < 10\GeV$ and $\abseta < 1.4$, the track resolutions were typically 1.5\% in \pt and 25--90 (45--150)\mum in the transverse (longitudinal) impact parameter \ipxy (\ipz)~\cite{tracking:CMS:2014pgm}. At the start of 2017, a new pixel detector was installed~\cite{Phase1Pixel}; the upgraded tracker measured particles up to $\abseta < 3.0$ with typical resolutions of 1.5\% in \pt and 20--75\mum in \ipxy for nonisolated particles of $1 < \pt < 10\GeV$~\cite{DP-2020-049}.

Physics events of interest are selected using a two-tiered trigger system. The first level, called the level-1 trigger, is composed of custom hardware processors and uses information from the calorimeters and muon detectors to select events at a rate of around 100\unit{kHz} within a fixed latency of about 4\mus~\cite{CMS:2020cmk}. The second level, known as the high-level trigger, consists of a farm of processors running a version of the full event reconstruction software optimized for fast processing, and reduces the event rate to around 1\unit{kHz} before data storage~\cite{CMS:2016ngn}.
A more detailed description of the CMS detector, together with a definition of the coordinate system used and the relevant kinematic variables, can be found in Ref.~\cite{CMS:2008xjf}.

\section{Event simulation}
\label{sec:mc}

Monte Carlo (MC) events are used to evaluate the signal acceptance, optimize selection criteria, and test the closure of the background estimation methods. 

The signal process is generated using the Hidden Valley module~\cite{Carloni:2010tw,Carloni:2011kk} in \PYTHIA version 8.240~\cite{Sjostrand:2014zea}, based on Ref.~\cite{Schwaller:2015gea}.
In the unflavored scenario, 
we choose the number of dark quark flavors to be $\Nfdark=7$, following Ref.~\cite{Schwaller:2015gea}. The running of the dark coupling constant with $Q^2$, where $Q$ is the momentum transfer, is faster for smaller \Nfdark, and the resulting showers have fewer dark mesons. 
In the flavor-aligned scenario, $\Nfdark=3$ is used, and \dcoupl is set to be diagonal, with all diagonal elements having the value \kappaZero.
For this scenario, the \PYTHIA Hidden Valley module is modified to produce the different dark hadron species at the desired occurrence frequencies based on the dark quark flavors.
For both cases, we consider a representation similar to QCD with $\Ncdark=3$. The dark quark masses are degenerate and equal to the confinement scale \Lamdark. The dark pion mass is set to be half of \Lamdark, and the dark $\rho$ meson mass four times the dark pion mass. 
These parameter choices are motivated by the assumptions made concerning dark matter energy density and energy scales in Ref.~\cite{Schwaller:2015gea}.
The natural width of $\dmed$ is set to 10\GeV, a relatively small value compared to the detector resolution. Under these assumptions, the free parameters of the model are limited to the mediator mass \mdmed, the dark pion mass \mdpi, and the dark pion lifetime \ctaudpi. Tables~\ref{tab:parameter_unflavored} and~\ref{tab:parameter_aligned} summarize the signal model parameters used in this search. The signal cross section, described in Section~\ref{sec:intro}, is computed at next-to-leading order (NLO), with the resummation of soft gluon emission included at next-to-leading-logarithmic accuracy. 

\begin{table}[htb]
    \centering
    \topcaption{Model parameters for the unflavored model.}
    \begin{tabular}{cc}    
        Model parameter & List of values \\ [\cmsTabSkip]
        \hline 
        \mdmed [\GeVns{}] & 1000, 1200, 1400, 1500, 1600, 1800, 2000, 2200, 2400, 2500 \\
        \mdpi [\GeVns{}] & 10, 20 \\
        \ctaudpi [mm] & 1, 2, 5, 25, 45, 60, 100, 150, 225, 300, 500, 1000 \\
    \end{tabular}
    \label{tab:parameter_unflavored}
\end{table}

\begin{table}[htb]
    \centering
\topcaption{Parameters used for the flavor-aligned model. 
In order to probe a range of lifetimes, the values of \kappaZero listed in columns 3--7 are tuned to give the desired \ctaumax values of 5, 25, 45, 100, and 500\mm.
In addition, samples were made with fixed $\kappaZero=1$, with a resultant value of \ctaumax that depends on the other model parameters.
}
    \begin{tabular}{ccccccc}
        \mdmed [\GeVns{}] & \mdpi [\GeVns{}] &
        \multicolumn{5}{c}{\kappaZero value}  \\ [\cmsTabSkip]
    \hline 
    \multirow{3}{*}{1000} &
              6 & 0.92 & 0.61 & 0.53 & 0.43 & 0.29 \\
          &  10 & 0.62 & 0.42 & 0.36 & 0.30 & 0.20 \\ 
          &  20 & 0.37 & 0.25 & 0.21 & 0.18 & 0.12 \\ 
    [\cmsTabSkip]    
    \multirow{3}{*}{1200}&
              6 & 1.10 & 0.73 & 0.63 & 0.52 & 0.35 \\
          &  10 & 0.75 & 0.50 & 0.43 & 0.35 & 0.24 \\
          &  20 & 0.45 & 0.30 & 0.26 & 0.21 & 0.14 \\
    [\cmsTabSkip]  
    \multirow{3}{*}{1400}&
              6 & 1.28 & 0.86 & 0.74 & 0.61 & 0.41 \\
          &  10 & 0.87 & 0.58 & 0.50 & 0.41 & 0.28 \\
          &  20 & 0.52 & 0.35 & 0.30 & 0.25 & 0.16 \\
    [\cmsTabSkip]  
    \multirow{3}{*}{1600}&
              6 & 1.47 & 0.98 & 0.85 & 0.69 & 0.46 \\
          &  10 & 1.00 & 0.67 & 0.58 & 0.47 & 0.32 \\
          &  20 & 0.59 & 0.40 & 0.34 & 0.28 & 0.19 \\
    [\cmsTabSkip]  
    \multirow{3}{*}{1800}&
              6 & 1.65 & 1.10 & 0.95 & 0.78 & 0.52 \\
          &  10 & 1.12 & 0.75 & 0.65 & 0.53 & 0.36 \\
          &  20 & 0.67 & 0.45 & 0.39 & 0.32 & 0.21 \\
    [\cmsTabSkip]  
    \multirow{3}{*}{2000}&
              6 & 1.83 & 1.23 & 1.06 & 0.87 & 0.58 \\
          &  10 & 1.25 & 0.84 & 0.72 & 0.59 & 0.40 \\
          &  20 & 0.74 & 0.50 & 0.43 & 0.35 & 0.23 \\
    [\cmsTabSkip]  
    \multirow{3}{*}{2200}&
              6 & 2.02 & 1.35 & 1.16 & 0.95 & 0.64 \\
          &  10 & 1.37 & 0.92 & 0.79 & 0.65 & 0.43 \\
          &  20 & 0.82 & 0.55 & 0.47 & 0.39 & 0.26 \\
    [\cmsTabSkip]  
    \multirow{3}{*}{2400}&
              6 & 2.20 & 1.47 & 1.27 & 1.04 & 0.70 \\
          &  10 & 1.50 & 1.00 & 0.87 & 0.71 & 0.47 \\
          &  20 & 0.89 & 0.60 & 0.51 & 0.42 & 0.28 \\
    [\cmsTabSkip]  
    \multirow{3}{*}{2500}&
              6 & 2.29 & 1.53 & 1.32 & 1.08 & 0.72 \\
          &  10 & 1.56 & 1.04 & 0.90 & 0.74 & 0.49 \\
          &  20 & 0.93 & 0.62 & 0.54 & 0.44 & 0.29 \\   
    \end{tabular}
    \label{tab:parameter_aligned}
\end{table}

\begin{sloppypar}We simulate SM QCD multijet events and \gjets events using the \MGvATNLO 2.6.5 event generator~\cite{Alwall:2014hca} at leading order with the MLM matching procedure~\cite{Alwall:2007fs}.\end{sloppypar}

\begin{sloppypar}In all cases parton showering and hadronization is performed using \PYTHIA{}8 with the CP5 underlying event tune~\cite{Sirunyan:2019dfx} and the NNPDF3.1 next-to-NLO parton distribution functions (PDFs)~\cite{Ball:2017nwa}. The response of the CMS detector is modeled using
\GEANTfour~\cite{Agostinelli:2002hh},
 and corrections are applied to the simulated samples to account for the differences in resolutions and efficiencies between the data and the simulation.\end{sloppypar}

\section{Event reconstruction and triggering}
\label{sec:reco}
A global ``particle-flow'' (PF) algorithm~\cite{CMS:2017yfk} aims to reconstruct all individual particles in an event, combining information provided by the tracker, calorimeters, and muon system. The reconstructed information from the different subsystems is used to build physics objects such as photons, leptons, jets, and missing transverse momentum~\cite{CMS:2018jrd,CMS:2016lmd,CMS:2019ctu}.

The pp interaction vertices are reconstructed by clustering tracks on the basis of their $z$ coordinates along the beamline at their points of closest approach to the beam axis using a deterministic annealing algorithm~\cite{Rose:1998dzq}. The position of each vertex is estimated with an adaptive vertex fit~\cite{Fruhwirth:2007hz}.

Multiple vertex candidates can be reconstructed because of additional pp interactions in a bunch crossing (pileup). 
The primary vertex (PV) is taken to be the vertex corresponding to the hardest scattering in the event, evaluated using tracking information alone, as described in Section 9.4.1 of Ref.~\cite{Contardo:2020886}. The contribution from charged-particle tracks from pileup interactions is reduced by rejecting those associated with other vertices~\cite{CMS:2020ebo}. The PV is required to be within 15\cm of the CMS detector center in the $z$ direction to ensure optimal reconstruction efficiency. 

Jets are reconstructed from the PF particles using the anti-\kt algorithm~\cite{Cacciari:2008gp,Cacciari:2011ma} with a distance parameter of 0.4. The jet momentum (energy) is calculated as the vectorial (scalar) sum of the momenta (energies) of all clustered particles. Corrections derived from data and simulation are applied to the jet energy. Jets that are consistent with the fragmentation of \PQb quarks (\PQb jets) are identified using the \DEEPJET discriminator~\cite{deepJet:Bols:2020bkb,BTV-16-002,CMS-DP-2023-005}. Both the medium and loose working points are used; the medium (loose) working point has an 85\% (90\%) probability of correctly identifying \PQb jets with $\pt>90\GeV$ and a 1\% (10\%) probability of misidentifying light-flavor jets as \PQb jets.
The scalar \PT sum of all jets within an event (\HT) is used to select energetic events.

The analysis considers data collected using triggers based on jet \PT and on \HT calculated from the summed \PT of online-reconstructed jets. The trigger used for the data collected in 2016 requires at least one jet with $\PT>450\GeV$ or $\HT>900\GeV$, while for 2017--2018, the \HT threshold is increased to 1050\GeV and there is no \PT requirement. The inclusion of a jet \PT trigger requirement for 2016 was necessary because, in part of the 2016 data taking period, some jets reaching the saturation energy for the level-1 trigger were mistakenly dropped from the \HT sum, resulting in a significant loss of efficiency. This loss was recovered by including the single-jet trigger requirement.
The \HT threshold was increased in 2017 to compensate for the higher instantaneous luminosity. The efficiencies for an event to pass any of these trigger conditions in both data and simulation are measured from data sets collected with an independent trigger that requires a muon with $\pt > 50\GeV$. In both cases, the efficiencies are close to unity above the signal selection \HT thresholds, and the difference
between the efficiencies as measured in simulation and data is applied as a correction to the simulation. 

\section{Analysis}
\label{sec:method}

The offline analysis selects energetic events with at least four jets.
Jets are classified as EJ candidates using information from tracks associated with the jets.
An overview of the event selection strategy is given in Section~\ref{sec:event}. Descriptions of the ``candidate'' EJ identification criteria (``tagging'') is given in Section~\ref{sec:emj_tag}. The algorithms used to optimize the selection criteria to maximize signal sensitivity, and the resulting event selection requirements are described in Section~\ref{sec:opt}.

\subsection{Event and physics object selection}
\label{sec:event}

While some SM hadrons, such as those containing \PQb quarks, have macroscopic decay lengths, tracks associated with SM jets mainly come from particles produced promptly (close to the collision primary vertex).
In EJs, when the lifetime of one or more of the dark mesons is long, a substantial fraction of the tracks can emerge from displaced vertices. 
In addition, in the flavor-aligned scenario, most of the dark pion decay products are \PQb quarks, which results in displaced tracks even when the dark pion decays immediately. We use as our displacement measure the \ipxy (\ipz), measured from the PV to the point of closest approach on the track trajectory, as this gives better sensitivity than reconstructing individual decay vertices. Jets used in this analysis are required to have $\PT>100\GeV$, as we are looking for EJs originating from a heavy-mediator decay, and $\abseta<2$, to ensure that they are well contained in the tracker. Jet candidates are also required to pass a set of quality criteria to reject spurious jets from instrumental sources~\cite{JETID_CMS-PAS-JME-16-003}. To reduce the probability of displaced tracks being incorrectly assigned to jets by the PF algorithm, high-purity tracks, as defined in Ref.~\cite{tracking:CMS:2014pgm}, with $\PT>1\GeV$ are associated with jets by requiring the angular separation between the jet direction and the track direction, $\dR=\sqrt{\smash[b]{(\deta)^2 + (\dphi)^2}} < \Rcut$, where \deta is the $\eta$ separation between the jet axis and the track, and \dphi is the separation in the azimuthal direction. We also reject tracks with $\abs{\ipz}>\ipzmax$ to suppress jets from pileup interactions. The values of \Rcut and \ipzmax are optimized for each signal model. 
If a track can be assigned to multiple jets, it is assigned to the jet with the smallest \dR separation. Jets are required to have at least one associated track so that the jet displacement measure can be calculated. 
To suppress events with a poorly reconstructed PV, we require that at least 10\% of associated tracks have a longitudinal displacement from the PV less than 0.01\cm.

Candidate signal events are required to have high \HT and at least four jets passing the criteria above. At least two of these four jets must be tagged as EJs. If more than four jets are found, the four jets with the largest \PT are used.  The \HT and EJ criteria are optimized for each signal model.

\subsection{EJ tagging}
\label{sec:emj_tag}

Two EJ tagging strategies are used in this analysis. The first selects candidate EJs via requirements on jet-level variables using track displacement measures (``model-agnostic selection'').
The second uses a graph neural network trained on specific signal models to determine an EJ tagging score (``machine learning or ML-based selection''). 
The model-agnostic method 
allows for a simpler reinterpretation of the results for alternate theoretical models not considered in this paper, while the ML-based approach achieves the best possible sensitivity for the specific models studied here. 

\begin{sloppypar}The model-agnostic EJ tagger uses different input features for the unflavored and flavor-aligned dark sector models, while the ML-based EJ tagger is trained separately for each class of model.\end{sloppypar}

\subsubsection{Model-agnostic EJ tagging}
\begin{figure}[htbp]
    \centering
    \includegraphics[width=0.48\textwidth]{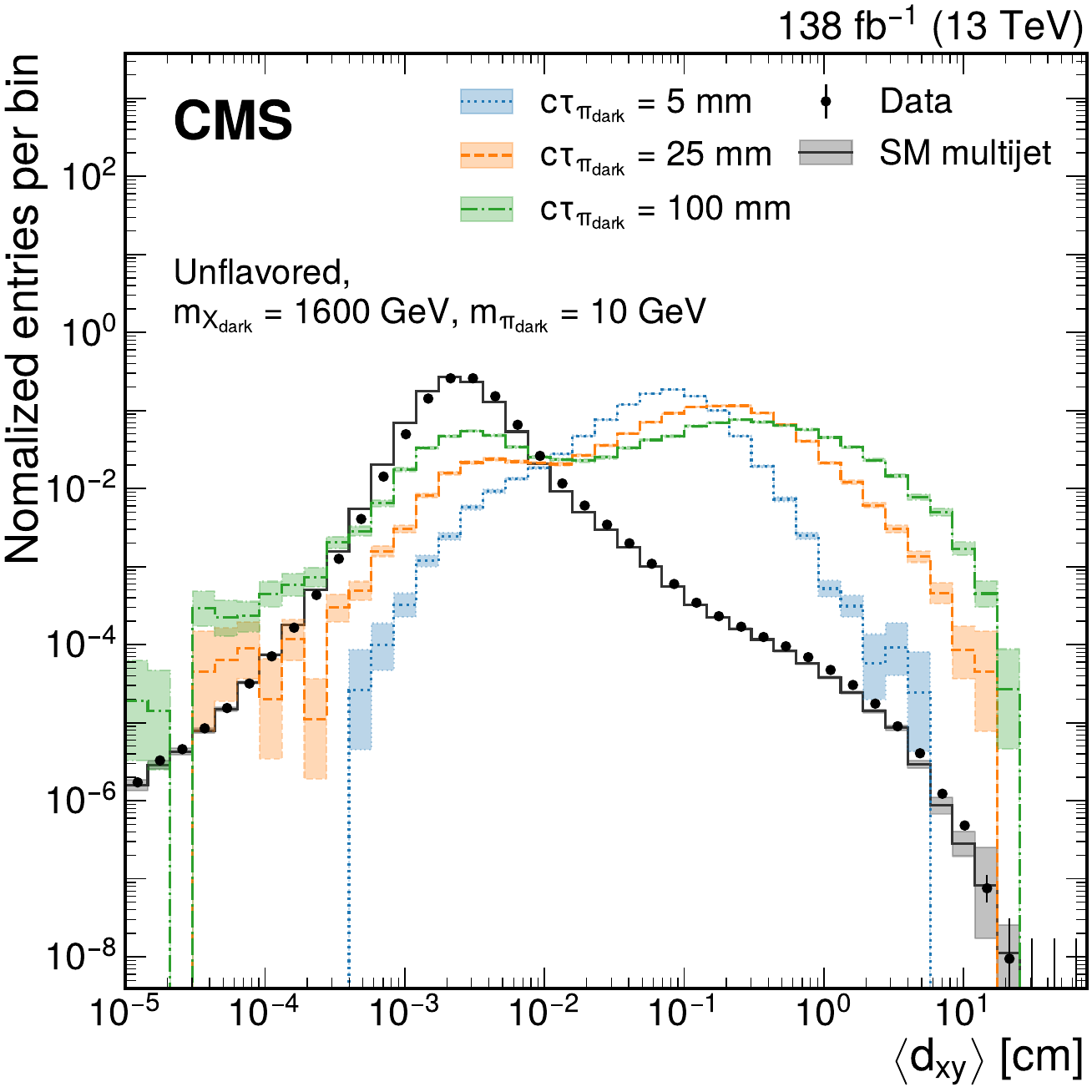}
    \hspace{0.02\textwidth}
    \includegraphics[width=0.48\textwidth]{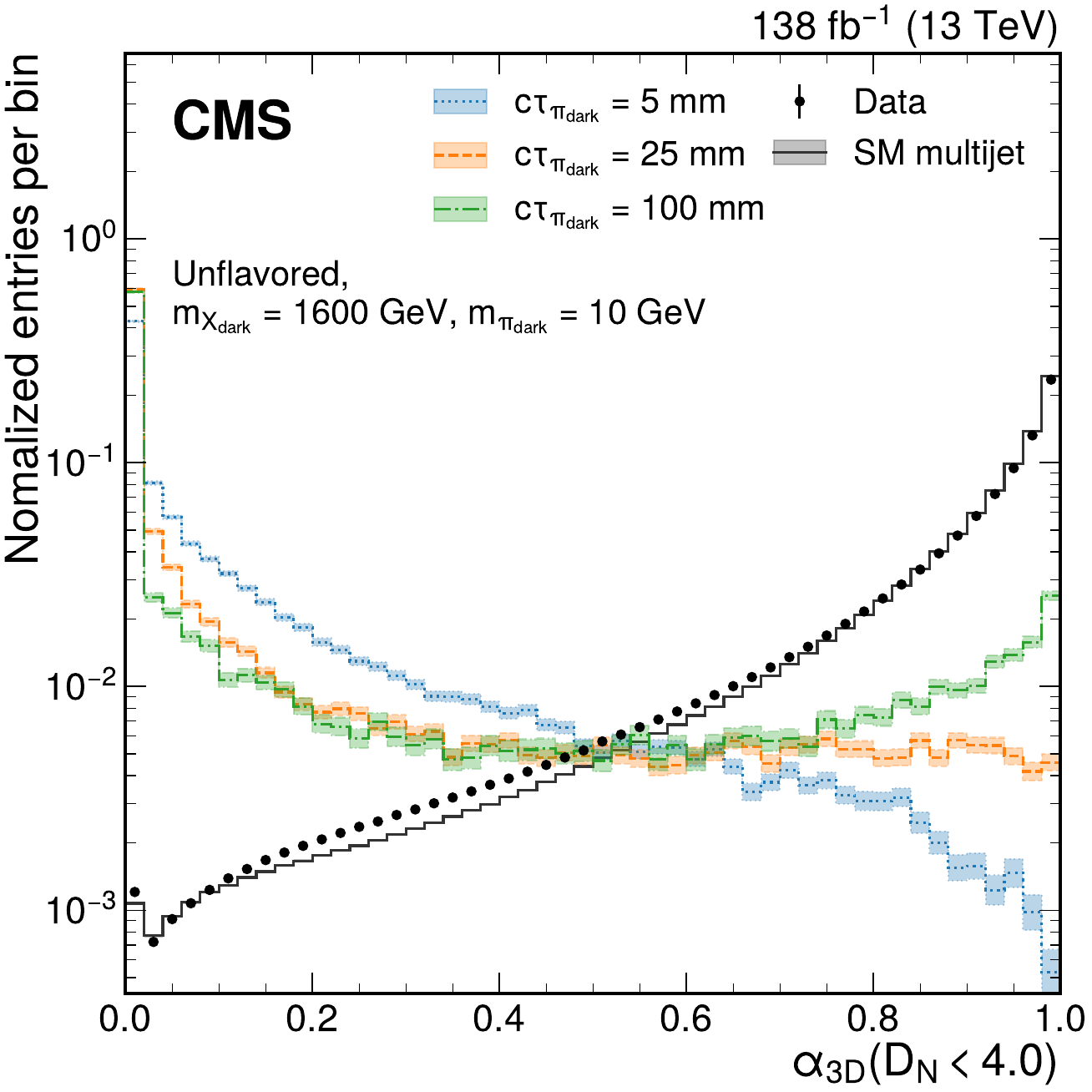}
    \caption{Distributions of the jet variables \medipxy (left) and \alphd with $\DNmax=4$ (right) used for the model-agnostic EJ tagging that targets the unflavored dark sector models are shown for data (points), SM multijet simulation (gray line), and signal jets in simulation (colored lines). The sums of the entries are normalized to unity.}
    \label{fig:cutbased-dr4}
\end{figure}

For the EJ tagging that targets the unflavored dark sector models,  $\Rcut=0.4$ is used. 
The following variables, which were also used in Ref.~\cite{CMSEMJ:Sirunyan:2018njd}, are used for the EJ tagging: 
\begin{itemize}
    \item \medipxy: this is the median \ipxy of tracks associated with a jet.
    \item \alphd: this is defined as
    \begin{cmsequation*}
        \alphd = \frac{\sum_{\DN < \DNmax} \pttrack}{\sum \pttrack},
    \end{cmsequation*}
    which is the ratio between the scalar \pt sum of the associated tracks with a pseudo-significance \DN smaller than the selection value \DNmax and the scalar \pt sum of all the associated tracks.   
    The pseudo-significance \DN is defined as:
    \begin{cmsequation*}
        \DN=\sqrt{\enc{\frac{\ipz}{0.01\cm}}^2+\enc{\dxysig}^2},
    \end{cmsequation*}
    where $\sigma(\ipxy)$ is the \ipxy uncertainty calculated from the covariance matrix of the fitted track trajectory. The \ipz significance is based on the PV $z$ resolution.
\end{itemize}
The distributions of these variables are presented in Fig.~\ref{fig:cutbased-dr4} for data events and for simulated signal and multijet background events.
The data and simulated events shown require $\HT > 1200\GeV$ and at least four jets with $\PT>100\GeV$ and $\abseta<2$, which is much less restrictive than the final signal selection requirements.

\begin{figure}[htbp]
    \centering
    \includegraphics[width=0.48\textwidth]{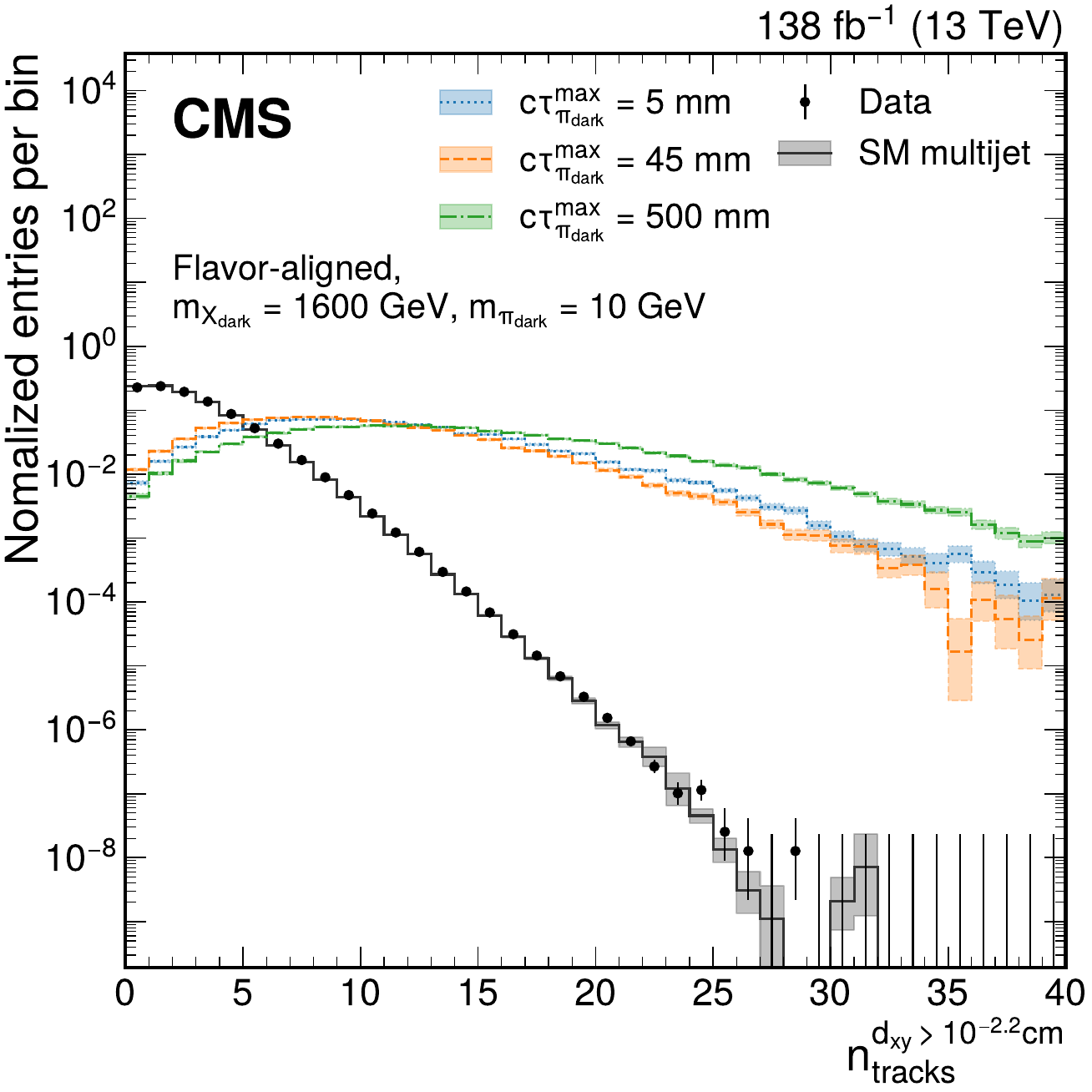}
    \hspace{0.02\textwidth}
    \includegraphics[width=0.48\textwidth]{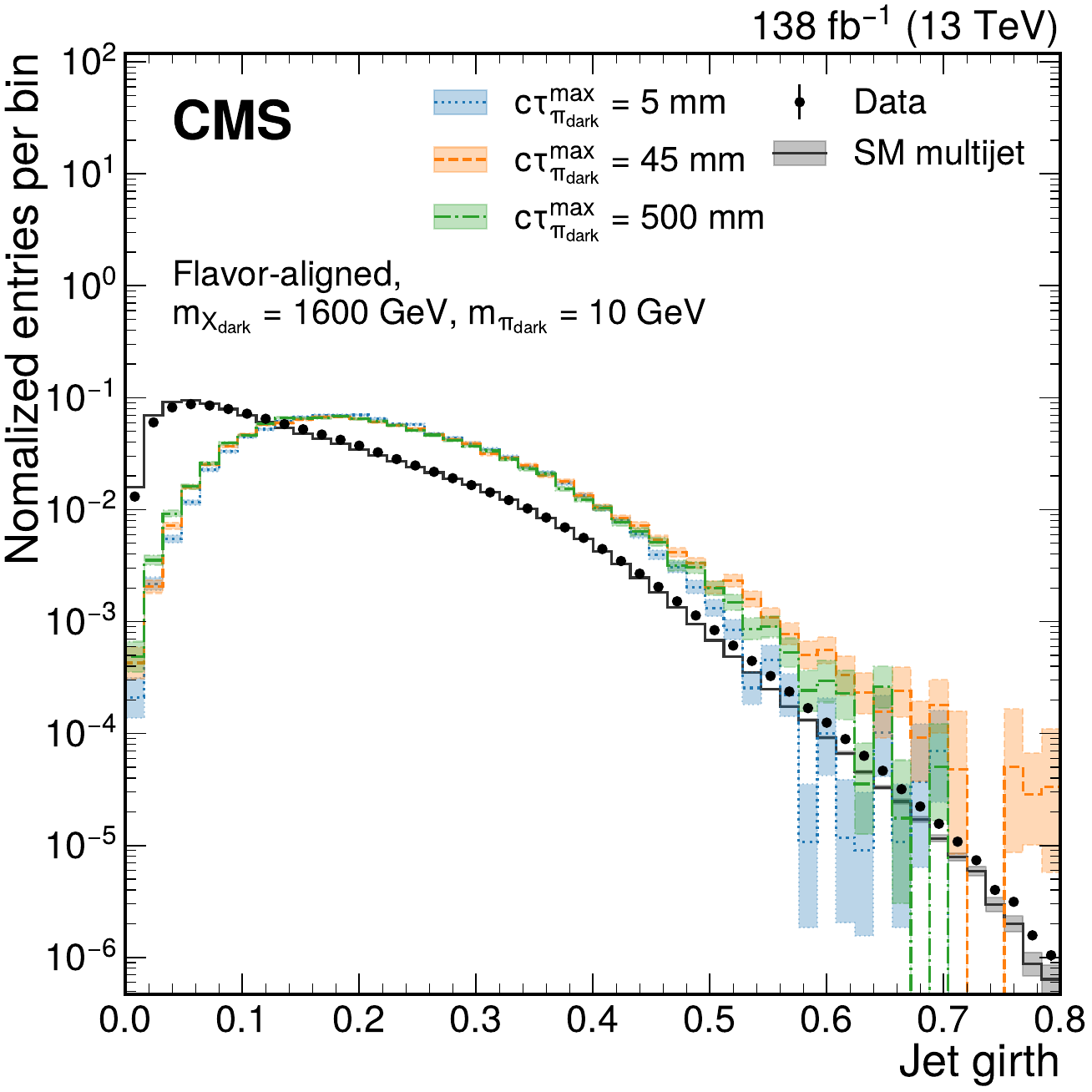}
    \caption{Distributions of the jet variables used for the model-agnostic EJ tagging targeting flavor-aligned dark sector models for jets obtained in data (points), SM multijet simulation (gray line), and simulated signal jets (colored lines). The distribution of the number of tracks with \mbox{$\trkdxy>10^{-2.2}\cm$} (jet girth) is shown on the left (right). The sums of the entries are normalized to unity.}
    \label{fig:cutbased-dr8}
\end{figure}

The flavor-aligned dark sector model has three different dark meson lifetime ranges:  long-lived dark mesons (up to 50\cm for the \kappaZero parameters considered in this search), dark mesons with a \PQb-hadron-like displacement from prompt dark meson decays into \PQb quarks, and dark mesons that decay promptly into light SM quarks.
In addition, since at least one dark quark in the flavor-aligned model couples to the \PQb quark, the jets often contain heavy-flavor hadrons, and the resulting EJ is wider than that of the unflavored model. Because of this, the tagging of EJs in the flavor-aligned model associates tracks with a jet using a proximity measure of $\Rcut=0.8$.
The selection for candidate flavor-aligned EJs is based on the following variables:
\begin{itemize}
    \item $\ntrack^{\trkdxy>\trkdxycmin}$, the number of tracks with \trkdxy greater than a threshold \trkdxycmin: this quantity exploits the tendency of the flavor-aligned dark sector to generate multiple long-lived SM mesons in the particle shower, resulting in a large number of displaced tracks.
    \item Jet girth: this is defined as the \pt-weighted \dR separation of tracks from the jet direction:
    \begin{cmsequation*}
        \text{Jet girth} = \frac{\sum_i \ptsup{i} \dR (i, \text{jet}) }{ \sum_i \ptsup{i}},
    \end{cmsequation*}
    where the index $i$ runs over all tracks associated with the jet of interest. This variable exploits the feature that particles in EJs tend to have a wider angular separation than SM jets because of the large mass of the dark mesons.
\end{itemize}
Figure~\ref{fig:cutbased-dr8} shows the distributions of the tagging variables used in the flavor-aligned analysis for data, simulated signal, and multijet background events. The data and simulated events follow the same requirements as those appearing in Fig.~\ref{fig:cutbased-dr4}.

As the jet-track association radius $\Rcut=0.8$ is larger than the $R=0.4$ value used to perform jet clustering, jets that have nearby soft jets can be misidentified as EJs.  To remove these candidates, a modified $N$-subjettiness variable \taunn is calculated as follows:
    \begin{enumerate}
        \item For each of the four candidate high-\PT jets, we consider all jets within $\dR<0.8$ of the selected jets that have $\PT>30\GeV$ in decreasing order of jet \PT. These will be used as the ``subjet'' collection to calculate \taunn. Using this definition, all candidate jets will have at least one subjet: the candidate jet itself.
        \item After determining the subjet collection assigned to each candidate jet, the computation of \taunn is then carried out up to the leading $n$ subjets:
    \begin{cmsequation*}
            \taun{n} = \frac{\sum_i \ptsup{i} \text{min}\lbrace\dR_{ij}\rbrace}{\sum_i 0.8 \ptsup{i}},
    \end{cmsequation*}
    similar to the definition of the original $N$-subjettiness~\cite{nsubjet:Thaler:2010tr}. The index $i$ runs over the tracks associated with the candidate jet, and $j$ runs from 1 up to $n$ for the subjet collection assigned to the selected jet.
    \end{enumerate}
The requirement that $\taun{2/1} = \taun{2}/\taun{1}>0.5$ is applied to all EJ candidates and is found to reliably suppress the misidentification of SM jets with nearby soft jets as EJs.

\subsubsection{The ML-based EJ tagging}

The ML-based tagger uses a graph neural network (GNN) based on  \textsc{ParticleNet}~\cite{ParticleNet} to directly incorporate the track information. Two separate GNNs are trained to classify EJs: one for the unflavored model (uGNN) and the other for the flavor-aligned model (aGNN). Each is trained, validated, and tested using all of the EJs from the signal samples and an equal number of background jets from background samples, weighted equally. The training, validation, and testing data subsets make up 60\%, 15\%, and 25\% of the full data set, respectively. The output of each GNN is a score ranging from 0 to 1, which is a measure of the probability that the jet is an EJ. Tracks are associated with a jet using $\Rcut=0.8$. To maximize the available information for the ML training, the tracks are not required to satisfy a \ipzmax requirement.  Each network is trained on all jets originating from a dark quark in the signal models of interest, and an equal number of jets taken from SM QCD simulation as the background sample.

The \deta and \dphi between each track and its associated jet are used as the track coordinate variables in the jet space. Each track in the jet is represented by a 5-feature vector containing:
\begin{itemize}
    \item $\dR(\text{track},\text{jet})$, as particles in EJs tend to have a wider angular separation than in the SM jets because of the heaviness of the dark mesons.
    \item $\ln(\pttrack/1\GeV)$, $\ln(\pttrack/\sum_i \PT^{i})$, as the combination of the dark shower and the decay of the mesons back to the SM sector causes the \PT of tracks to be smaller on average for EJs than for SM jets.
    \item $T(\ipxy)$, $T(\ipz)$. The transformation function $T(x)$ is applied to the track displacement variables, to reduce the range of values input to the GNN while preserving the variables' sign and continuity. It is defined as:
        \begin{equation*}
            T(x) = \sign(x)\ln\left(\left\vert \frac{x}{1\cm}\right\vert +1\right).
        \end{equation*}
    This transformation was found to give comparable or better performance than a standard scaling.
\end{itemize}

\begin{figure}[htbp]
    \centering
    \includegraphics[width=0.48\textwidth]{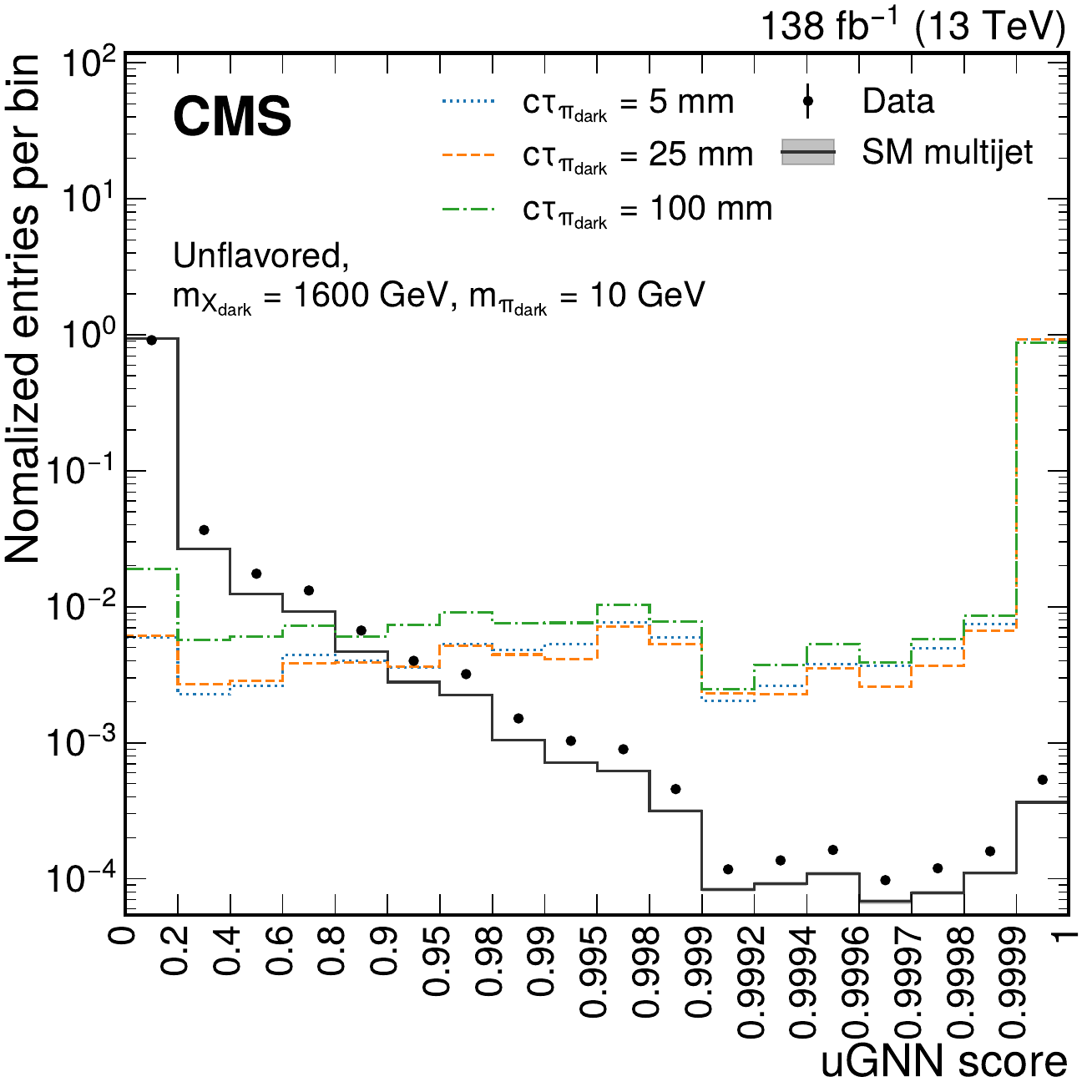}
    \hspace{0.02\textwidth}
    \includegraphics[width=0.48\textwidth]{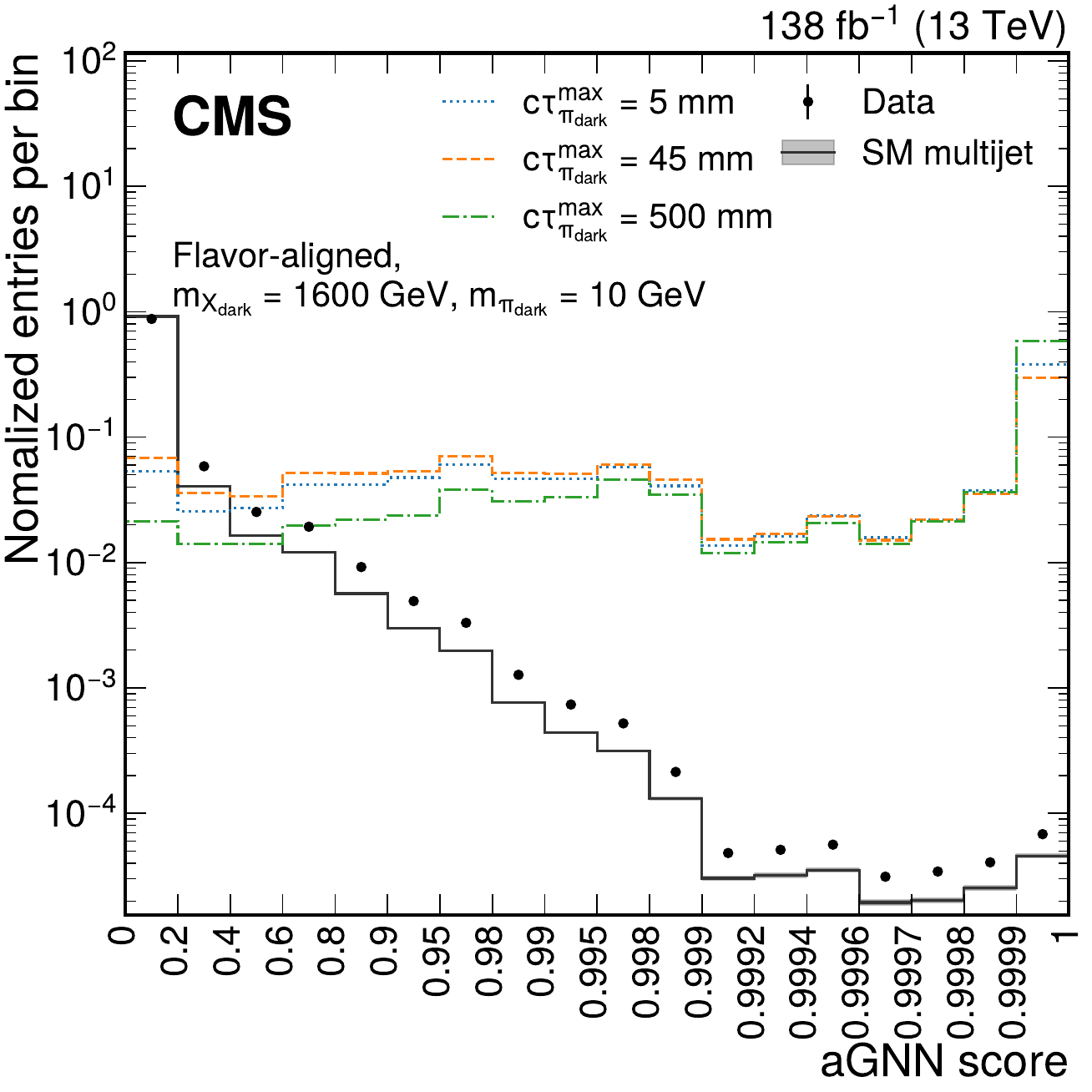}
    \caption{Distributions of the output score of the uGNN (left) and aGNN (right) for the data (points with error bars), SM multijet simulation (dark gray line), and signal simulation (colored lines).  The signal distributions in the left (right) plot are generated from the unflavored (flavor-aligned) model. Bins are chosen to correspond to the jet selection criteria defined in Table~\ref{tab:emj_tagger_gnn}. The uncertainties in the SM multijet simulation are too small to be visible. The systematic uncertainties in the simulated signal distributions are small and have been omitted for reasons of clarity. The sums of the entries are normalized to unity.}
    \label{fig:gnn_score}
\end{figure}

The impact parameter $\ipxy$ is the most influential feature. 
Figure~\ref{fig:gnn_score} shows the output score distributions for signal and background, demonstrating good separation.  The signal distributions are similar despite the wide range of dark meson \ctau values.

\subsection{Determining the EJ tagging and event selection criteria}
\label{sec:opt}

Searches for each of the signal models listed in Tables~\ref{tab:parameter_unflavored} and~\ref{tab:parameter_aligned} are conducted. An optimization procedure is performed for each signal model to determine the best threshold on the event \HT, jet \PT, and the EJ tagging variables by maximizing \sigmaopt, defined as:
\begin{cmsequation}
    \sigmaopt = \frac{S}{\sqrt{S+B+\beta^2B^2}},
    \label{eq:sigopt}
\end{cmsequation}
where $S$ ($B$) is the number of simulated signal (background multijet) events passing the selection thresholds and $\beta$ is the estimated relative systematic uncertainty in the background, taken to be 10\%. 
To reduce the number of sets of selection criteria,
similar selection sets are grouped, and a representative set that gives at least 90\% of the original best significance value is used for all the models in the group. 
Additional selection sets with higher background selection efficiencies are used to validate the background estimation calculations.  Their selection criteria are detailed in Section~\ref{sec:bkg}.

The selection criteria used for EJ tagging are shown in Tables~\ref{tab:emj_tagger_dr4}--\ref{tab:emj_tagger_gnn}, and those for the event selection in Table~\ref{tab:emj_cutset}.
In the model-agnostic approach, longer signal model lifetimes in the unflavored scenario tend to favor selection criteria with larger \medipxy selection thresholds. On the other hand, in the flavor-aligned scenario, longer signal model lifetimes favor selection criteria with higher \trkdxycmin thresholds. In the ML-based approach, GNN EJ taggers generally exhibit consistent performance across varying signal model lifetimes, and the signal models with larger mediator particle masses generally are targeted by selection criteria with higher \HT threshold. The specific selection criteria employed for each EJ model, along with the signal selection efficiency for each model under various selection criteria, are documented in the HEPdata record~\cite{HEPData}.
For the unflavored model, at $\ctaudpi=45\mm$, the model-agnostic taggers yield a maximum signal selection efficiency of $\approx$40\%, while the ML-based taggers yield a maximum signal selection efficiency of $\approx$60\%.
The signal selection efficiency of the model-agnostic taggers drops to a few percent for very short-lived ($\approx$1\mm) dark pions, and the efficiency of both taggers drops to a few percent for very long-lived ($\approx$1000\mm) dark pions.
For the flavor-aligned model, at $\ctaumax=45\mm$, the model-agnostic taggers yield a maximum signal selection efficiency of $\approx$25\%, while the ML-based taggers yield a maximum signal selection efficiency of $\approx$40\%.
This efficiency remains fairly stable along the full maximum proper lifetime range for both taggers.

\begin{table}[htb]
    \centering
    \topcaption{Emerging jet selection criteria for the model-agnostic analysis designed for the unflavored scenario. The validation regions are discussed in Section~\ref{sec:bkg}. The symbols in parentheses indicate a minimum $(>)$ or maximum $(<)$ requirement.}
    \label{tab:emj_tagger_dr4}
    \begin{tabular}{ccccc}
    
    Tag name 
    & $\trkdz$ [cm] $(<)$
    & $\med{\ipxy}$ [cm] $(>)$
    & $\DN$ $(<)$ 
    & $\alphd$ $(<)$ \\ [\cmsTabSkip]
    \hline
    u-tag 1          & 0.5  & $10^{-1.6}$ & 4.0  & 0.25  \\
    u-tag 2          & 1.0  & $10^{-1.4}$ & 8.0  & 0.25 \\
    u-tag 3          & 5.0  & $10^{-1.2}$ & 8.0  & 0.25 \\
    u-tag 4          & 5.0  & $10^{-1.2}$ & 12.0 & 0.15 \\
    u-tag 5          & 5.0  & $10^{-1.0}$ & 12.0 & 0.15 \\
    validation u-tag & 0.5  & $10^{-1.6}$ & 4.0  & 0.40 \\
    
    \end{tabular}
\end{table}

\begin{table}[htb]
    \centering
    \topcaption{Emerging jet selection criteria for the model-agnostic analysis designed for the flavor-aligned scenario. The validation tag is described in Section~\ref{sec:bkg}.  The symbols in parentheses indicate a minimum $(>)$ or maximum $(<)$ requirement.}
    \label{tab:emj_tagger_dr8}
    \begin{tabular}{ccccc}
    Tag name & 
    \trkdz [cm] $(<)$ &
    $\trkdxycmin$ [cm] &
    $\ntrack^{\trkdxy>\trkdxycmin}$ $(>)$ &
    Jet girth $(>)$ \\ [\cmsTabSkip]
    \hline
    a-tag 1          & 0.5  & $10^{-2.2}$ & 12  & 0.05  \\
    a-tag 2          & 0.5  & $10^{-2.2}$ & 12  & 0.1  \\
    a-tag 3          & 0.5  & $10^{-2.3}$ & 14  & 0.0  \\
    a-tag 4          & 0.5  & $10^{-2.4}$ & 14  & 0.1  \\
    validation a-tag & 0.5  & $10^{-2.4}$ & 12  & 0.0  \\
    
    \end{tabular}
\end{table}

\begin{table}[htb]
    \centering
    \topcaption{The GNN score range used to identify a jet as an EJ. The uGNN (aGNN) tag indicates that the tagger uses the output score of the GNN trained on the unflavored (flavor-aligned) simulated signal samples. The validation tags are described in Section~\ref{sec:bkg}.}
    \label{tab:emj_tagger_gnn}
    \begin{tabular}{ccc}

    Tag name  & Score min. & Score max.\\ [\cmsTabSkip]
    \hline
    uGNN tag 1          & 0.9997 & 1\hphantom{.0000} \\ 
    uGNN tag 2          & 0.9998 & 1\hphantom{.0000} \\ 
    uGNN tag 3          & 0.9996 & 1\hphantom{.0000} \\
    uGNN validation tag & 0.998\hphantom{0} & 0.9995 \\
    [\cmsTabSkip]
    aGNN tag 1           & 0.9953 & 1\hphantom{.0000} \\ 
    aGNN tag 2           & 0.9993 & 1\hphantom{.0000} \\ 
    aGNN tag 3           & 0.9983 & 1\hphantom{.0000} \\
    aGNN validation tag  & 0.99\hphantom{00} & 0.995\hphantom{0}\\

    \end{tabular}
\end{table}

\begin{table}[htb]
    \centering
    \topcaption{Event selection criteria used for the analysis. The validation selection criteria are described in Section~\ref{sec:bkg}.}
    \label{tab:emj_cutset}
    \begin{tabular}{ccccccc}
    Selection set                              &
    \HT [\GeVns{}]                                 &
    \multicolumn{4}{c}{Jet \pt [\GeVns{}] ($>$)} &
    EJ tagger \\ [\cmsTabSkip]
    \hline
    u-set 1                                   & $>$1600    & 275 & 250 & 250 & 150 & u-tag 1             \\
    u-set 2                                   & $>$1600    & 200 & 200 & 150 & 150 & u-tag 2             \\
    u-set 3                                   & $>$1600    & 200 & 150 & 100 & 100 & u-tag 3             \\
    u-set 4                                   & $>$1500    & 200 & 150 & 100 & 100 & u-tag 4             \\
    u-set 5                                   & $>$1200    & 200 & 150 & 100 & 100 & u-tag 5             \\
    u-set validation                          & 1000--1200 & 100 & 100 & 100 & 100 & validation u-tag    \\
    [\cmsTabSkip]
    a-set 1                                   & $>$1500    & 200 & 150 & 100 & 100 & a-tag 1             \\
    a-set 2                                   & $>$1800    & 250 & 250 & 200 & 200 & a-tag 2             \\
    a-set 3                                   & $>$1200    & 275 & 250 & 250 & 200 & a-tag 2             \\
    a-set 4                                   & $>$1500    & 275 & 250 & 250 & 100 & a-tag 3             \\
    a-set 5                                   & $>$1800    & 200 & 150 & 100 & 100 & a-tag 4             \\
    a-set validation                          & 1000--1200 & 100 & 100 & 100 & 100 & validation a-tag    \\
    [\cmsTabSkip]
    uGNN set 1                                & $>$1350    & 170 & 120 & 120 & 100 & uGNN tag 1          \\
    uGNN set 2                                & $>$1750    & 300 & 260 & 250 & 250 & uGNN tag 2          \\
    uGNN set 3                                & $>$1800    & 240 & 180 & 180 & 100 & uGNN tag 3          \\
    uGNN validation                           & $>$1000    & 100 & 100 & 100 & 100 & uGNN validation tag \\
    [\cmsTabSkip]
    aGNN set 1                                & $>$1300    & 200 & 140 & 120 & 100 & aGNN tag 1          \\
    aGNN set 2                                & $>$1650    & 300 & 250 & 200 & 200 & aGNN tag 2          \\
    aGNN set 3                                & $>$1400    & 270 & 220 & 220 & 120 & aGNN tag 3          \\
    aGNN validation                           & $>$1000    & 100 & 100 & 100 & 100 & aGNN validation tag \\
    \end{tabular}
\end{table}

\section{Background estimation}
\label{sec:bkg}

The signal region (SR) for this analysis is constructed from events with two or more tagged EJs. The main source of background for this analysis is the production of four SM jets, where two or more
of these jets are misidentified as EJs. We estimate the number of SM events passing each selection criteria in the SR by constructing a control region (CR) with identical event and jet kinematic requirements, but where exactly one jet is tagged as an EJ. We estimate the fraction of signal events in the CR to be no more than $10^{-5}$. 
The probability of an SM jet being misidentified as an EJ (mistag rate) is heavily dependent on the underlying jet flavor, as SM jets containing \PQb hadrons also have displaced tracks. We therefore estimate the misidentification probability \epsf as a function of the underlying jet flavor $f$, where $f$ is \PQb for \PQb jets and \cPq for other jet flavors, and the jet \PT.

When the probability of a jet being misidentified as an EJ is characterized and parameterized using \PT, flavor fraction, and jet type, the number of background events in the SR is estimated using counts in the CR and the mistag rate according to the following formula:
\begin{cmsequation}
    \label{eq:scalefactor_sum}
    N_{\text{SR}} = 
       \sum_{\text{events}\,\in\,\text{CR}} 
    \frac{
       \frac{1}{2!}\enc{\sum_{i} \epsi \prod_{j\neq i} (1-\epsj)}
     + \frac{1}{3!}\enc{\sum_{i\neq j}\epsi\epsj \prod_{k\neq i,j} (1-\epsk)}
     + \frac{1}{4!}\enc{\sum_{i\neq j\neq k}\epsi\epsj\epsk}
    }{\prod_{i} (1-\epsi)},
\end{cmsequation}
where \epsi is an abbreviation of $\epsilon(f_i, \ptsup{i})$, the estimated mistag rate of jet $i$ in each event. The jet indices in the summations and products run over all jets that are not EJ tagged. The denominator calculates the fraction of events that have exactly one EJ-tagged jet, while the three terms in the numerator calculate the fractions of events with two, three, and four EJ-tagged jets, respectively. The common factor for the estimated mistag rate for the EJ-tagged jet exists in both the numerator and denominator, and therefore does not appear in the expression. The factorials in the numerator are needed to correct for repeated counting in the unordered sum over the jet indices. For example, the first term in the numerator needs a factor 2! to correctly handle events where the two highest \PT jets are tagged as EJs and the next two jets are not tagged; without this factor, the unordered sum would count these events twice. A detailed derivation of the equation is given in Ref.~\cite{Savard:2897463}. Because the underlying flavor of each jet is difficult to determine, we approximate \epsi in Eq.(\ref{eq:scalefactor_sum}) using a flavor-fraction-averaged misidentification rate $\epsavg(\PT)$ and the estimated \PQb jet fraction \Fb{CR} of all non-EJ-tagged jets in the CR:
\begin{cmsequation}
    \label{eq:mistag_avg}
    \epsavg(\PT) = \Fb{CR} \epsb + \left(1-\Fb{CR}\right) \epsq.
\end{cmsequation}

\begin{figure}[htbp]
    \centering
    \includegraphics[width=0.48\textwidth]{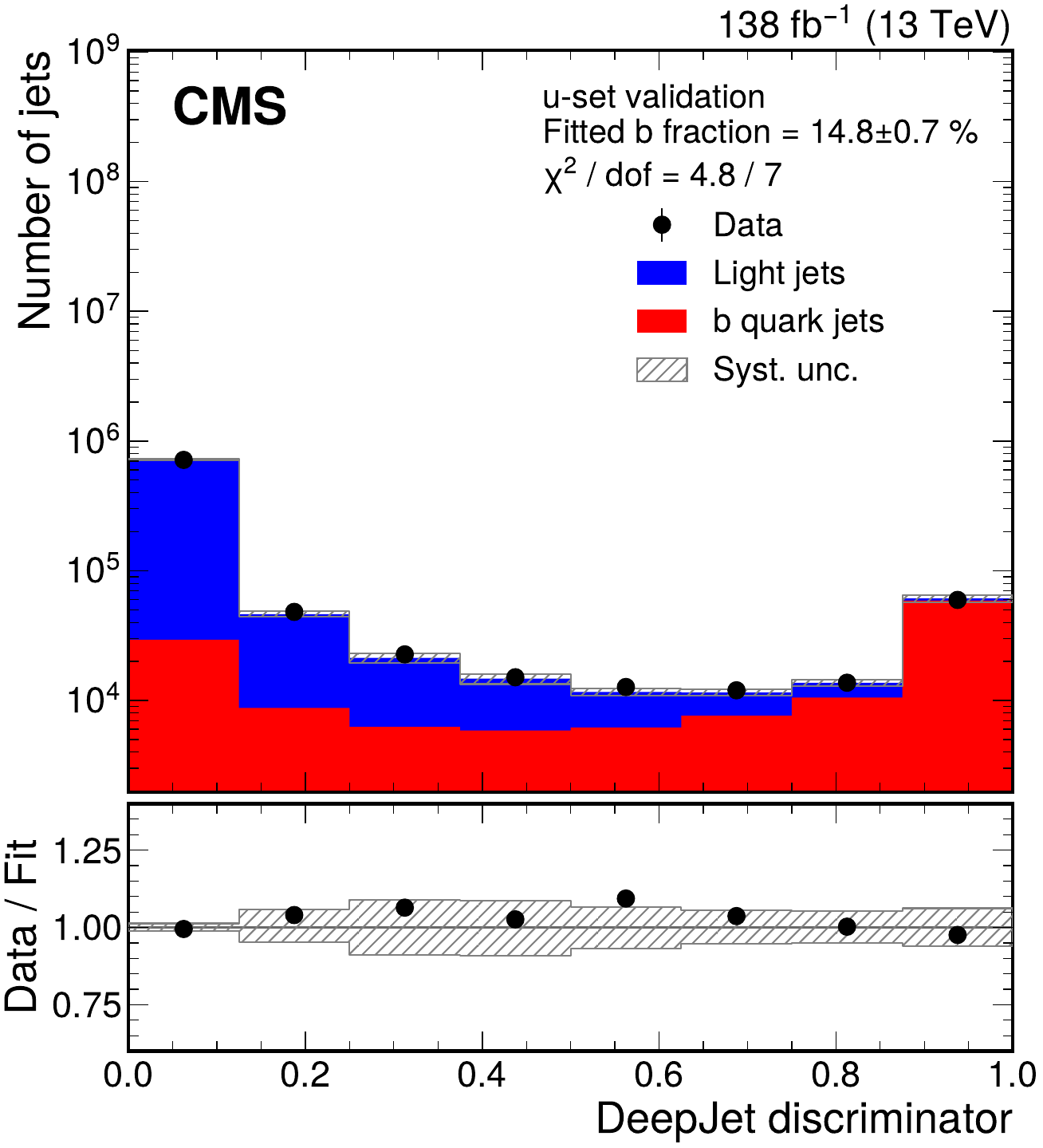}
    \hspace{0.02\textwidth}
    \includegraphics[width=0.48\textwidth]{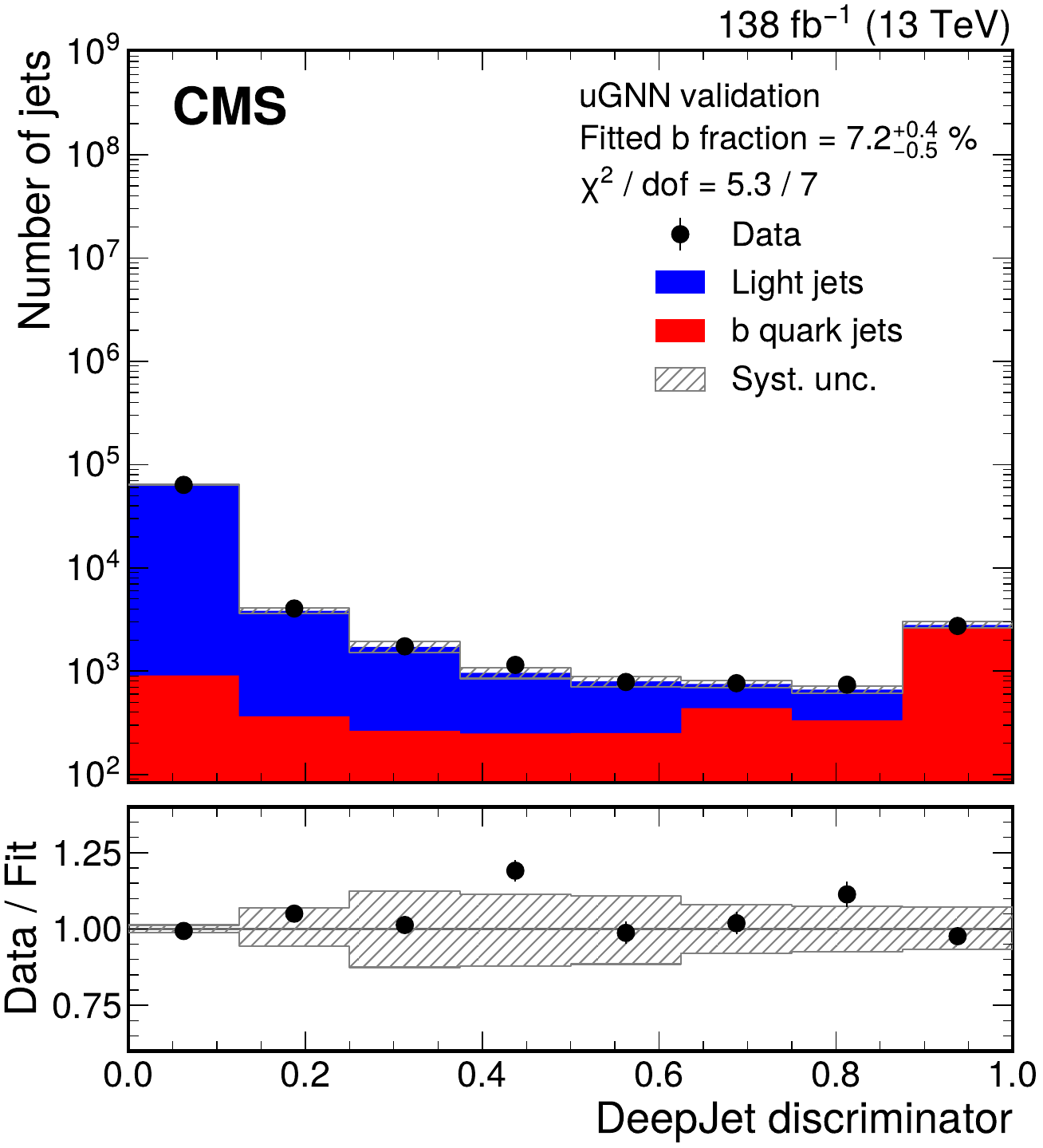}
    \caption{Template fit of the \DEEPJET discriminator used to determine the \PQb jet fraction of the non-EJ tagged jets for data events that pass the ``u-set validation'' (uGNN validation) selection criteria shown on the left (right), except with the requirement on the number of EJ-tagged jets  changed from 2 to 1. The lower panels show the ratio of the number of jets in the data compared to the sum of the fitted template distributions.}
    \label{fig:bfrac_fit}
\end{figure}

The value of $\Fb{CR}$ is estimated by fitting the CMS \DEEPJET discriminator~\cite{deepJet:Bols:2020bkb} spectrum of the non-EJ-tagged jets in the CR to two template distributions obtained from SM multijet MC events.  One template contains the discriminator value for jets identified using generator-level information as containing a \PQb quark, and the complement template is used for the other jets. The template distributions are varied within the measured uncertainties~\cite{BTV-16-002,CMS-DP-2023-005}. The normalization factor of the \PQb quark template, $\Fb{CR}$, is the only free parameter. The fit is performed for each of the selection criteria listed in Table~\ref{tab:emj_cutset}. Figure~\ref{fig:bfrac_fit} gives an example of the fit performance for the ``u-set validation'' and the ``uGNN validation'' selection criteria, defined in Table~\ref{tab:emj_cutset}.

\begin{figure}[htb]
    \centering
    \includegraphics[width=0.48\textwidth]{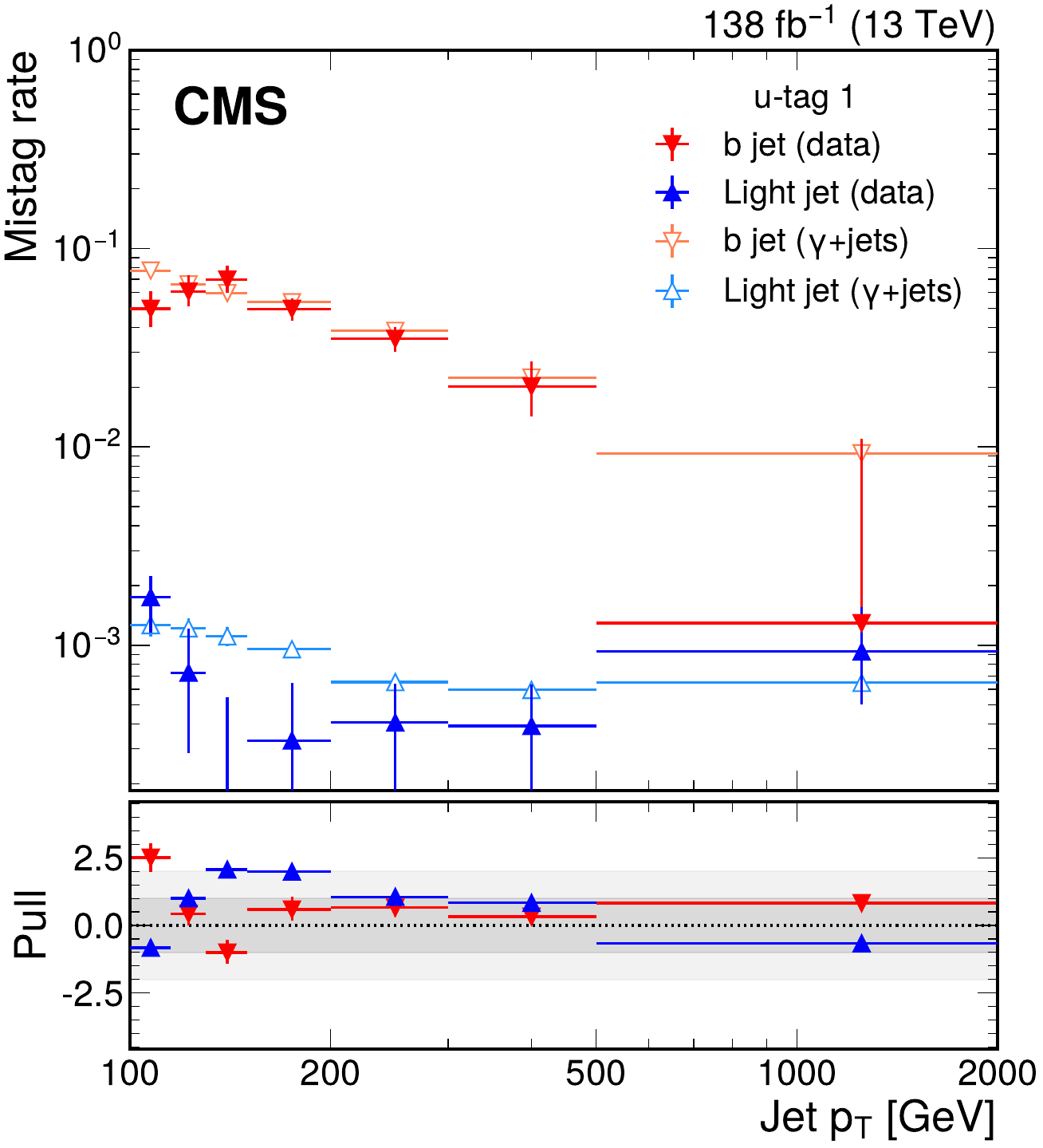}
    \hspace{0.02\textwidth}
    \includegraphics[width=0.48\textwidth]{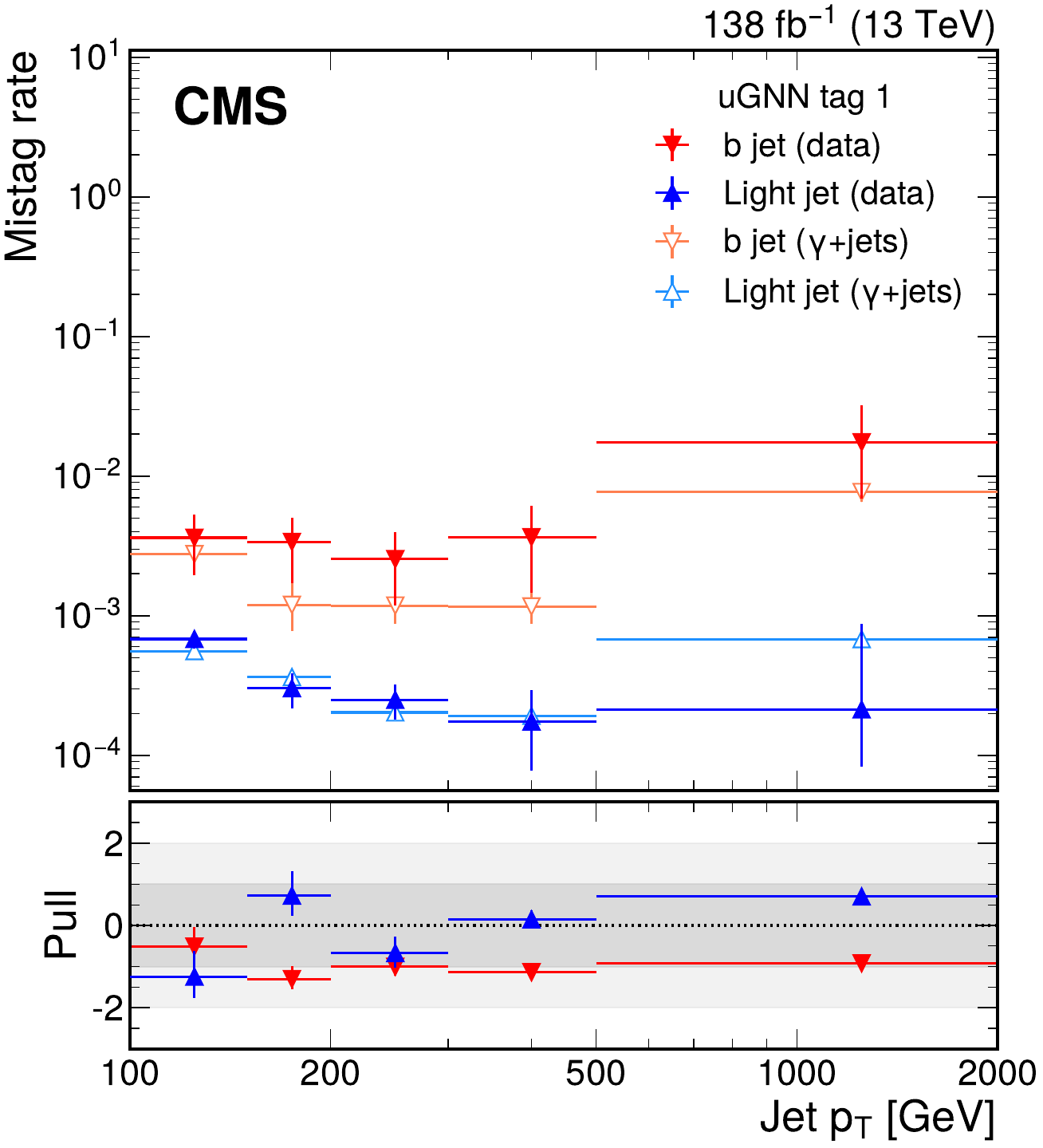}
    \caption{The EJ tagger misidentification probability for \PQb quark jets (red, orange) and light jets (light blue, dark blue) as a function of jet \PT for the model-agnostic tagger ``u-tag 1'' (left) and the ML-based tagger ``uGNN tag 1'' (right), as defined in Tables~\ref{tab:emj_tagger_dr4} and~\ref{tab:emj_tagger_gnn}, evaluated using data (red, dark blue) and generator-level flavor information from simulated samples (orange, light blue) in events containing a high-\pt photon. 
    The lower panel shows the pull, defined as the difference between the mistag rate calculated in simulation and mistag rate measured in data, scaled down by the uncertainty measured in data. The error bars indicate the uncertainties in the mistag rates measured in simulation scaled by the uncertainties measured in data.
    }
    \label{fig:mistag_rate}
\end{figure}

The evaluation of the mistag rate is performed in a signal-free region (FR)
that consists of events passing a high-\PT photon trigger and containing an isolated photon in the ECAL barrel with $\PT>200\GeV$. The FR events are required to have either one or two jets passing the jet selection criteria described in Section~\ref{sec:event}. To obtain the misidentification probability for different jet flavors, we further divide the FR into a \PQb-enriched region (ER) and a \PQb-depleted region (DR) by imposing \PQb tagging requirements on an additional jet with $\PT>50\GeV$ and $\abseta<2.4$ that passes a set of noise-rejection criteria~\cite{JETID_CMS-PAS-JME-16-003}. As a significant fraction of heavy-flavored jets originate from gluon splitting in this event sample, \PQb tagging requirements on this extra jet will change the \PQb jet fraction of the selected jets. Events are classified as ER if the additional jet passes \DEEPJET \PQb tagging at the medium working point, and are classified as DR if the additional jet fails \DEEPJET \PQb tagging at the loose working point.
Assuming that the ER and DR have the same mistag probabilities for the selected jets, and the only difference between these regions is the overall \PQb jet fraction, the misidentification rate for all the selected jets in a specific region $\text{XR}$, $\epsilon^{\text{XR}}(\PT)$, can be expressed as a linear combination of \epsb and \epsq:
\begin{cmsequation}
\label{eq:flavored_mis}
\begin{split}
    \epsilon^{\text{ER}}(\PT) &= \Fb{ER}(\PT) \epsb + \left(1-\Fb{ER}(\PT)\right)\epsq, \\
    \epsilon^{\text{DR}}(\PT) &= \Fb{DR}(\PT) \epsb + \left(1-\Fb{DR}(\PT)\right)\epsq,
\end{split}
\end{cmsequation}
where $\Fb{XR}(\PT)$ is the estimated \PQb jet fraction of region $\text{XR}$ in bins of jet \PT.
The \PQb jet fraction is obtained by fitting the \DEEPJET \PQb discriminator distribution to templates obtained using jets from a simulated \gjets sample. On average, the selected jets in the ER and DR have \PQb jet fractions of 15\% and 4\%, respectively. The linear relation in Eq.~\eqref{eq:flavored_mis} can then be inverted to obtain the mistag rates for different underlying jet flavors \epsf, which is used in Eq.~\eqref{eq:mistag_avg} to estimate the misidentification rate of jets in the CR. Figure~\ref{fig:mistag_rate} gives examples of the estimated EJ tagger misidentification probabilities for ``u-tag 1'' and ``uGNN tag 1'' taggers, defined in Tables~\ref{tab:emj_tagger_dr4} and~\ref{tab:emj_tagger_gnn}.

Because the Phase 1 upgrade of the pixel detector~\cite{Phase1Pixel} improved the track reconstruction performance, the evaluation of the expected background is performed separately for the pixel Phase 0 geometry (2016) and Phase 1 geometry (2017--2018). 

\subsection{Validation tests}
To validate the background estimation by checking the soundness of the calculations, closure tests are performed in validation regions (VRs) that use selection criteria that are orthogonal to the SRs. 
The VRs are defined using signal-like selection criteria that require at least two jets passing a validation EJ tagger. 
The observed number of events in the VR is compared to the number predicted using our background estimation technique.  For the model-agnostic EJ tagging VR, the \HT requirement of the validation event selection is inverted to ensure orthogonality with the SRs and small signal contamination.  The EJ identification requirements for the validation EJ tagger are based on the u-tag 1 and a-tag 1 requirements for unflavored and flavor-aligned models, respectively, with one jet-variable selection requirement relaxed to further reduce the effect of any small signal contamination. The VR for the ML-based EJ tagging uses a validation tagger that selects jets  with GNN scores in a region disjoint from the SRs. To further increase the number of events that pass the ML-based VR, we relax the \HT and jet \PT requirements. The full list of the selection criteria used for the VRs is given in Tables~\ref{tab:emj_tagger_dr4}--\ref{tab:emj_cutset}. The signal contamination in these VRs is less than 1\% for all surveyed signal models.

The number of events passing the VR selection criteria with at least two jets passing the validation jet tag, as well as the predicted number using the estimation described in Eq.~\eqref{eq:scalefactor_sum}, is given in Table~\ref{tab:result_closure}. No significant deviation between the observed and the estimation results is observed, indicating that the background estimation calculation is robust.

\begin{table}[htb]
    \topcaption{The observed yield of events in data satisfying the validation selection criteria with at least two jets passing the corresponding validation tag, and the estimation based on the misidentification rate calculated using validation events with exactly one jet passing the validation tagger scaled by the factor given in  Eq.~\eqref{eq:scalefactor_sum}. The statistical and systematic uncertainties are reported for the estimated yields.}
    \label{tab:result_closure}
    \renewcommand{\arraystretch}{1.4}

    \centering
    \begin{tabular}{cZO}
        \multicolumn{1}{c}{Selection set} &
        \multicolumn{2}{r@{\symmdelim}}{\TabHeaderA} &
        \multicolumn{2}{@{\hspace{0em}}c@{\symmdelim}}{\TabHeaderB} &
        \multicolumn{2}{@{\hspace{0em}}l}{\TabHeaderC} &
        \multicolumn{1}{c}{\TabHeaderO} \\ [\cmsTabSkip]
        \hline
        \TabHeadSpace
        u-set validation   &  1220&   & $\leftscript{80}{80}$ & $\leftscript{}{}$    &  210&  &  1484   \\
        a-set validation   &    77&  & $\leftscript{ 4}{ 4}$  & $\leftscript{}{}$    &   15&  &   118   \\
        uGNN validation    &    29&.3 & $\leftscript{ 3}{ 7}$ & $\leftscript{.4}{.0}$ &   8&.2 &    21   \\
        aGNN validation    &    29&.1 & $\leftscript{ 2}{ 6}$ & $\leftscript{.6}{.1}$ &   5&.2 &    37   \\
    \end{tabular}    
\end{table}

\section{Systematic uncertainties}
\label{sec:unc}

\subsection{Background uncertainties} 
The main sources of systematic uncertainty in this search arise from the background estimation method based on control samples in data. 
The kinematic differences between the SM jets in the CR photon-triggered data used for the misidentification rate calculation and the SM jets in the \HT-triggered data used in the SR may lead to slightly different misidentification rates. We estimate the corresponding uncertainties by applying the background prediction method in Eq.~\eqref{eq:scalefactor_sum} to simulated SM multijets events, using either the mistag rate from the SM multijet simulation or from the \gjets simulation. The uncertainty is taken as the difference in the background predictions obtained with the two mistag rates. 
There is also an uncertainty in the flavor composition used in the misidentification rate estimations. 
This uncertainty is estimated from simulated background events by comparing the flavor-decomposition estimate described in Section~\ref{sec:bkg} with one derived from generator-level flavor information. 
Finally, there is the uncertainty associated with the choice of variables used to parameterize the mistag rate. 
This uncertainty is estimated by comparing the estimation results when parameterizing the mistag rate as a function of jet \pt versus track multiplicity, and as a function of coarse versus fine \pt binning, as the mistag rate has a significant dependence on the parameter chosen and binning scheme used.
The uncertainties in the flavor composition and mistag rate parameterization are larger on average than the uncertainty due to CR and SR differences. The model-agnostic method is more affected by changes in the chosen mistag rate parameter because the jet-level variables used depend on track multiplicity. The GNN method is more sensitive to the mistag rate binning scheme because of statistical fluctuations due to the small number of jets in some of the bins. The contribution of all the background uncertainty sources for various signal model search regions is summarized in Table~\ref{tab:bkg_unc_summary}.
These three uncertainty sources are considered to be uncorrelated and summed in quadrature. The resulting uncertainties are given in Tables~\ref{tab:result_closure} and~\ref{tab:eventyield}.
The difference between the observed and estimated yields in the a-set validation, shown in Table~\ref{tab:result_closure}, was investigated. No anomaly was found in the data, and thus it is assumed to arise from a statistical fluctuation. 

\begin{table}[htb]
    \centering
    \topcaption{Mean and standard deviation (std.) of the relative uncertainty calculated on the background estimations,  by source, in percent.}
    \label{tab:bkg_unc_summary}
    \begin{tabular}{crrrrrrrr}
    & \multicolumn{4}{c}{Model-agnostic taggers} & \multicolumn{4}{c}{ML-based taggers} \\
    [\cmsTabSkip] 
    Uncertainty source  & 
    \multicolumn{2}{c}{Unflavored} &   
    \multicolumn{2}{c}{Flavor-aligned} & 
    \multicolumn{2}{c}{Unflavored} &   
    \multicolumn{2}{c}{Flavor-aligned}\\
    & mean & std. & mean & std. & mean & std. & mean & std.\\ [\cmsTabSkip]
    \hline 
    Parameterization            &   27.0  &  7.3 &   11.8 &  0.9 &  8.3  &  3.3 &   44.6 &  9.9 \\ 
    CR/VR differences           &    6.4  &  2.4 &   11.1 &  5.4 & 16.4  &  9.1 &   12.3 &  2.9 \\ 
    Jet flavor evaluation       &   10.7  &  2.2 &   24.5 &  6.3 & 10.5  &  1.8 &   15.5 &  6.5 \\
    \end{tabular}
\end{table}

\subsection{Signal uncertainties}
Various sources of uncertainty affecting the signal yields are also considered. The integrated luminosity uncertainties for the 2016, 2017, and 2018 data-taking periods are 1.2, 2.3, and 2.5\%, respectively~\cite{CMS:2021xjt, CMS-PAS-LUM-17-004, CMS-PAS-LUM-18-002}.
There is a small difference in the trigger efficiency between data and simulation, with a ratio varying between 0.95 and 1.00. A correction factor compensates for this difference, and its statistical uncertainty  is propagated to an uncertainty in the signal acceptance. The evaluation of jet energy correction uncertainties is performed as a function of jet \PT and $\eta$,  propagated to all jet-related kinematic variables, and then to an uncertainty in the signal acceptance. Uncertainty sources  are treated as fully correlated across the years for the jet energy scale (JES), and uncorrelated for the jet energy resolution (JER). While the pileup distribution in the simulation is reweighted to match the data, there is a 4.6\% uncertainty in the total inelastic cross section~\cite{Sirunyan:2018nqx}, which is treated as fully correlated across the years. 
The impact parameter distribution is slightly broader in the data CRs than in simulation.  A smearing function that corrects this is applied to tracks in the simulated samples, and the resulting changes to the signal acceptances are used to estimate the modeling uncertainty. The corrected values are also used to recompute the GNN score. The uncertainty in the acceptance from the uncertainties in the PDFs is evaluated by reweighting events with different PDF variations~\cite{Ball:2013hta,Ball:2017nwa}. 
As reconstruction efficiency differences between data and simulation for displaced tracks are difficult to evaluate, we varied the reconstruction efficiency for tracks with $\ipxy>4\cm$ (approximately the distance to the second layer of the pixel detector) by 10\% and found an ${\approx}1\%$ overall change in signal acceptance. As the measured difference in reconstruction efficiency in data and MC is expected to be much smaller~\cite{CMS-DP-2020-035, CMS:2022fut}, no uncertainty is assigned.
The uncertainty from the factorization scale (\muF) and renormalization scale (\muR) choices is estimated by independently varying \muF and \muR by factors of 2 and 0.5~\cite{Cacciari:2003fi,Catani:2003zt}. 
A summary of the estimated uncertainties for the various signal model acceptances is given in Table~\ref{tab:sig_unc_summary}.

\begin{table}[htb]
    \centering
    \topcaption{Mean and standard deviation (std.) of the relative uncertainty calculated on the unflavored and flavor-aligned samples,  by source, in percent.}
    \label{tab:sig_unc_summary}
    \begin{tabular}{crrrrrrrr}
    & \multicolumn{4}{c}{Model-agnostic taggers} & \multicolumn{4}{c}{ML-based taggers} \\
    [\cmsTabSkip] 
    Uncertainty source  & 
    \multicolumn{2}{c}{Unflavored} &   
    \multicolumn{2}{c}{Flavor-aligned} & 
    \multicolumn{2}{c}{Unflavored} &   
    \multicolumn{2}{c}{Flavor-aligned}\\
    & mean & std. & mean & std. & mean & std. & mean & std.\\ [\cmsTabSkip]
    \hline 
    Integrated luminosity           &    1.8  &  0.6 &    1.8 &  0.6 & 1.8  &  0.6 &    1.8 &  0.6 \\ 
    Trigger efficiency   &    0.3  &  0.1 &    0.3 &  0.1 & 0.3  &  0.1 &    0.3 &  0.1 \\ 
    JES                  &    1.0  &  1.3 &    0.8 &  0.7 & 1.3  &  0.9 &    0.7 &  0.4 \\ 
    JER                  &    0.3  &  0.4 &    0.3 &  0.3 & 0.2  &  0.3 &    0.2 &  0.1 \\
    Pileup reweighting  &     1.6  &  1.4 &    1.4 &  1.2 & 0.9  &  0.8 &    1.0 &  0.9 \\ 
    Track modeling in sim.    &    0.2  &  0.3 &    1.4 &  1.8 & 0.3  &  0.8 &    0.5 &  0.6 \\ 
    PDF        & ${<}0.1$  &  ${<}0.1$ & ${<}0.1$ &  ${<}0.1$ & ${<}0.1$  &  $<0.1$ & ${<}0.1$ &  ${<}0.1$\\
    \muF, \muR & ${<}0.1$  &  ${<}0.1$ & ${<}0.1$ &  ${<}0.1$ & ${<}0.1$  &  ${<}0.1$ & ${<}0.1$ &  ${<}0.1$\\   
    \end{tabular}
\end{table}

\section{Results}
\label{sec:results}

\begin{figure}[tb]
    \centering
    \includegraphics[width=0.48\textwidth]{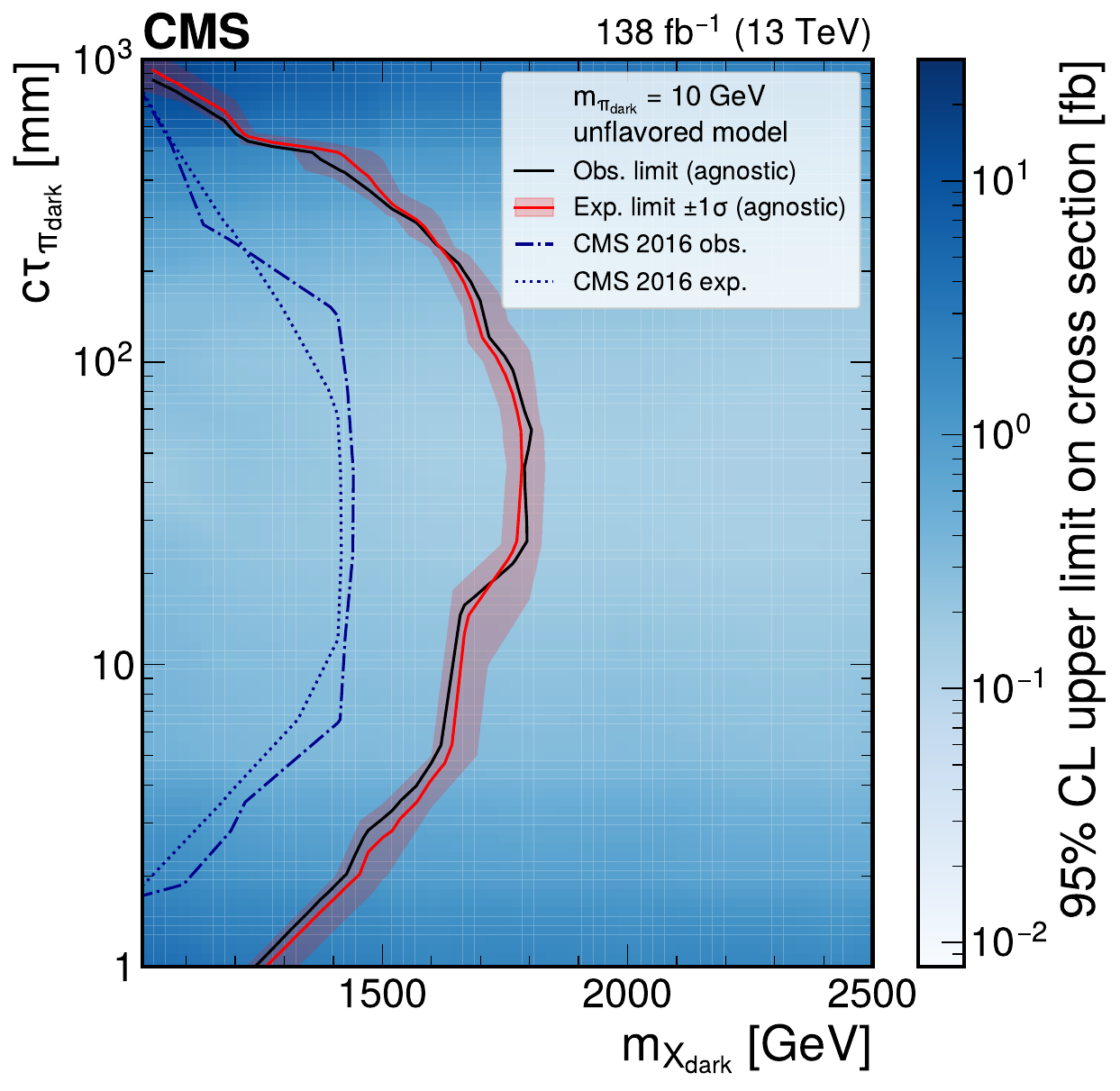}
    \hspace{0.02\textwidth}
    \includegraphics[width=0.48\textwidth]{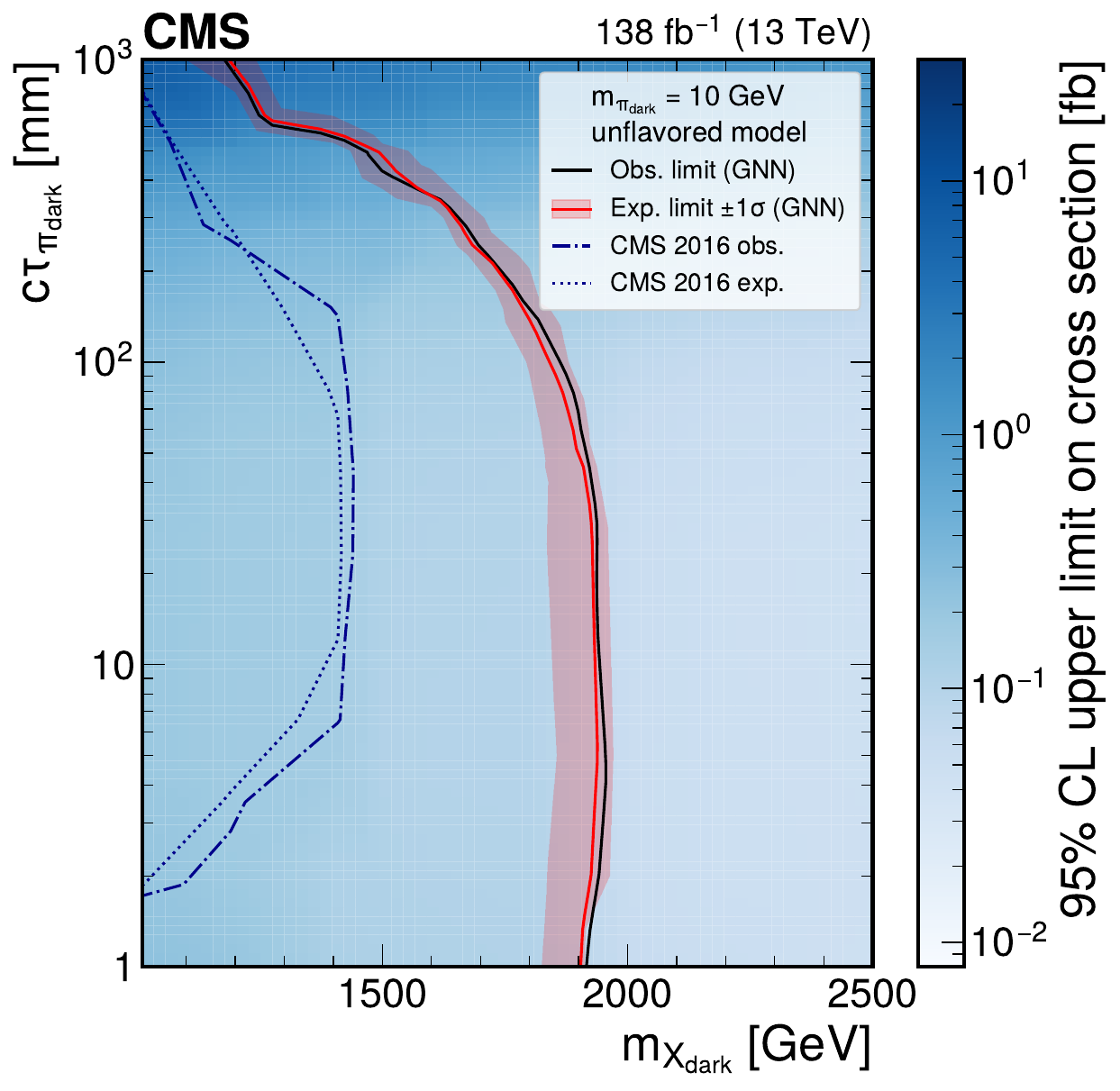} \\
    \includegraphics[width=0.48\textwidth]{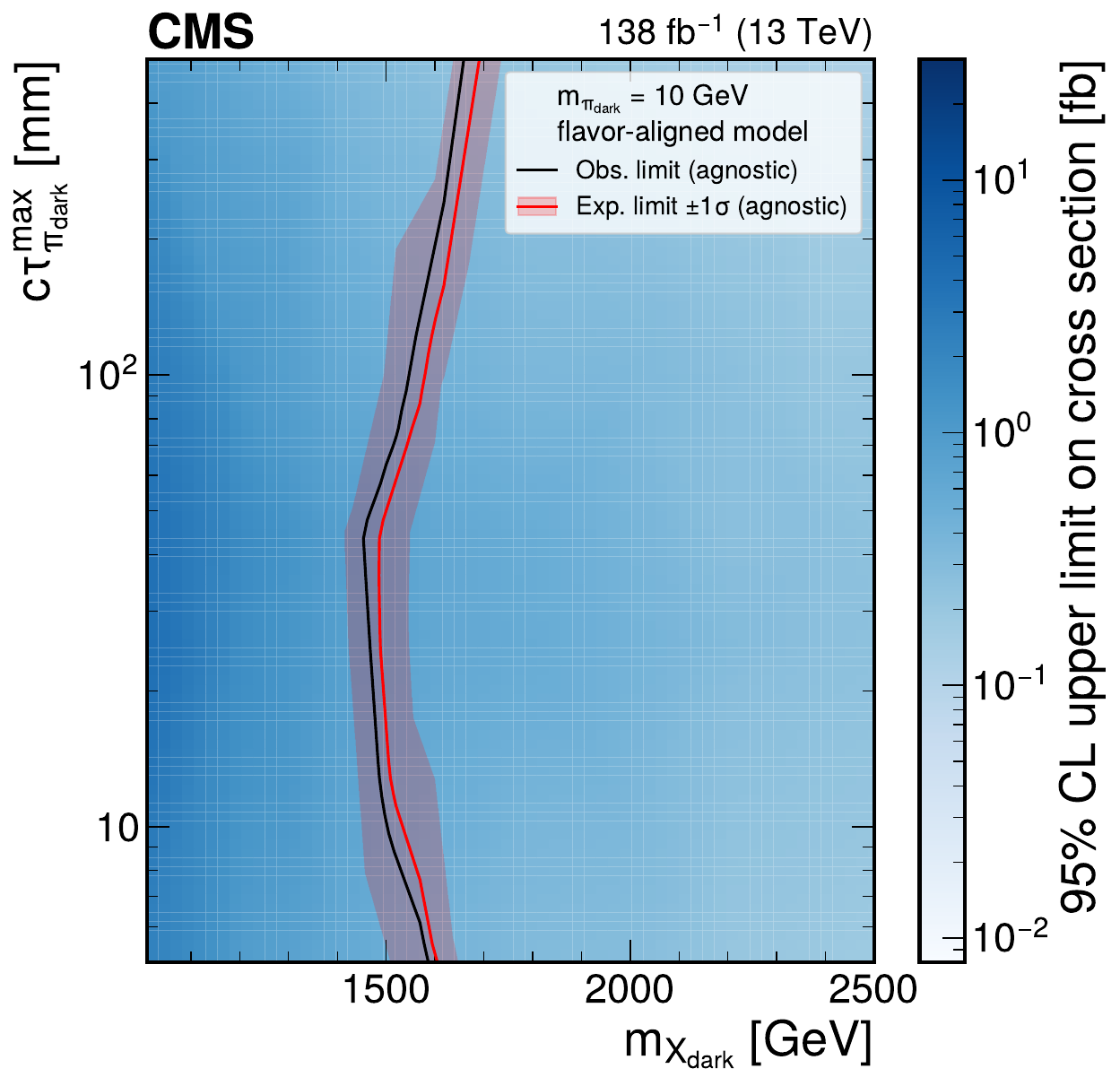}
    \hspace{0.02\textwidth}
    \includegraphics[width=0.48\textwidth]{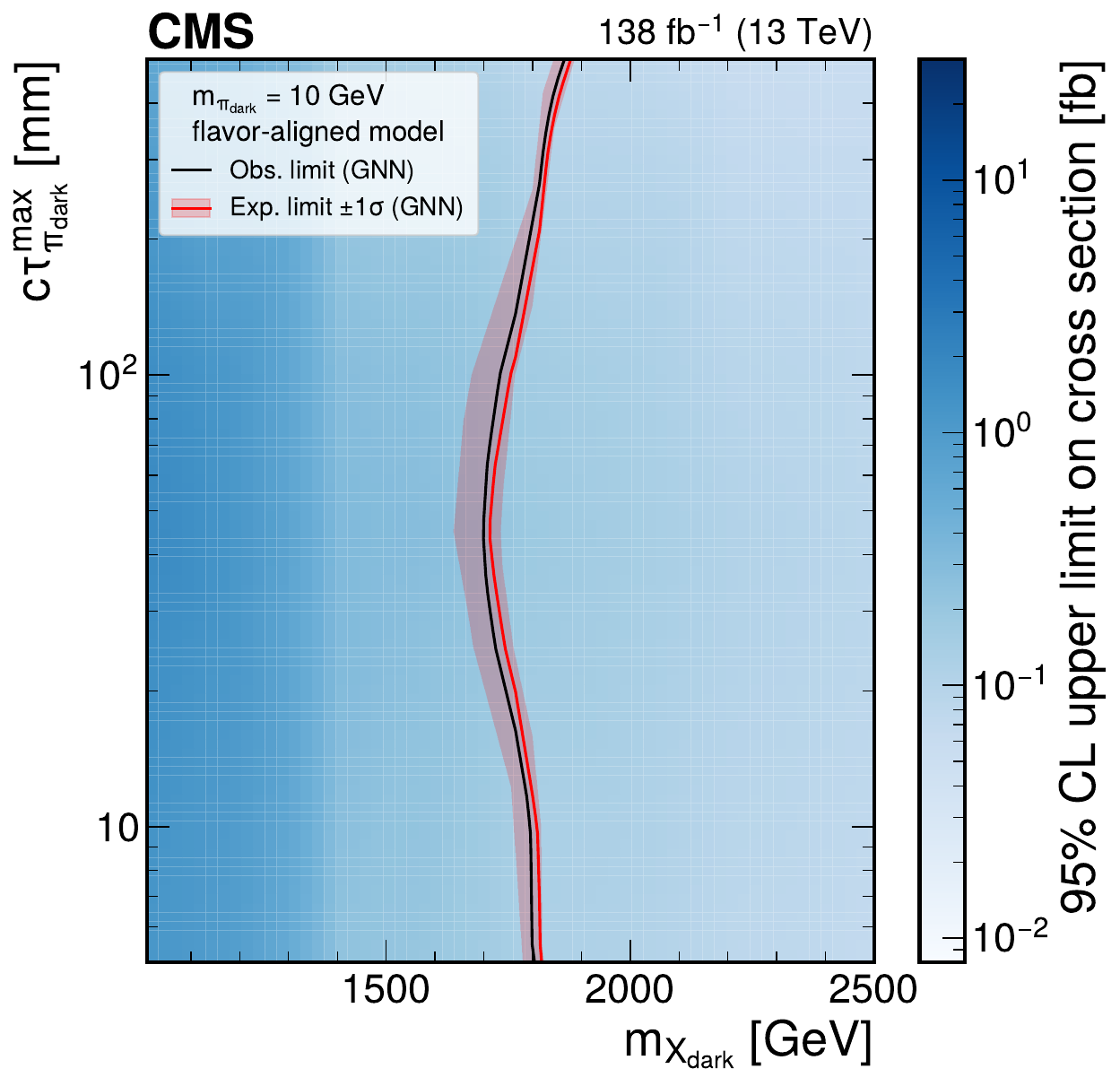}
    \caption{The 95\% \CL upper limits on the production cross section for various signal models in the unflavored scenario (upper plots) and the flavor-aligned scenario (lower plots) with $\mdpi=10\GeV$ using the model-agnostic (GNN) EJ tagging method, on the left (right). The red curve is the expected exclusion limit, with the band representing its 68\% \CL variation. The black curve is the observed limit. The dark blue dotted curves in the upper plots are the expected and observed limits previously obtained by CMS~\cite{CMSEMJ:Sirunyan:2018njd}.}
    \label{fig:result_mpi10}
\end{figure}

\begin{figure}[tbh]
    \centering
    \includegraphics[width=0.48\textwidth]{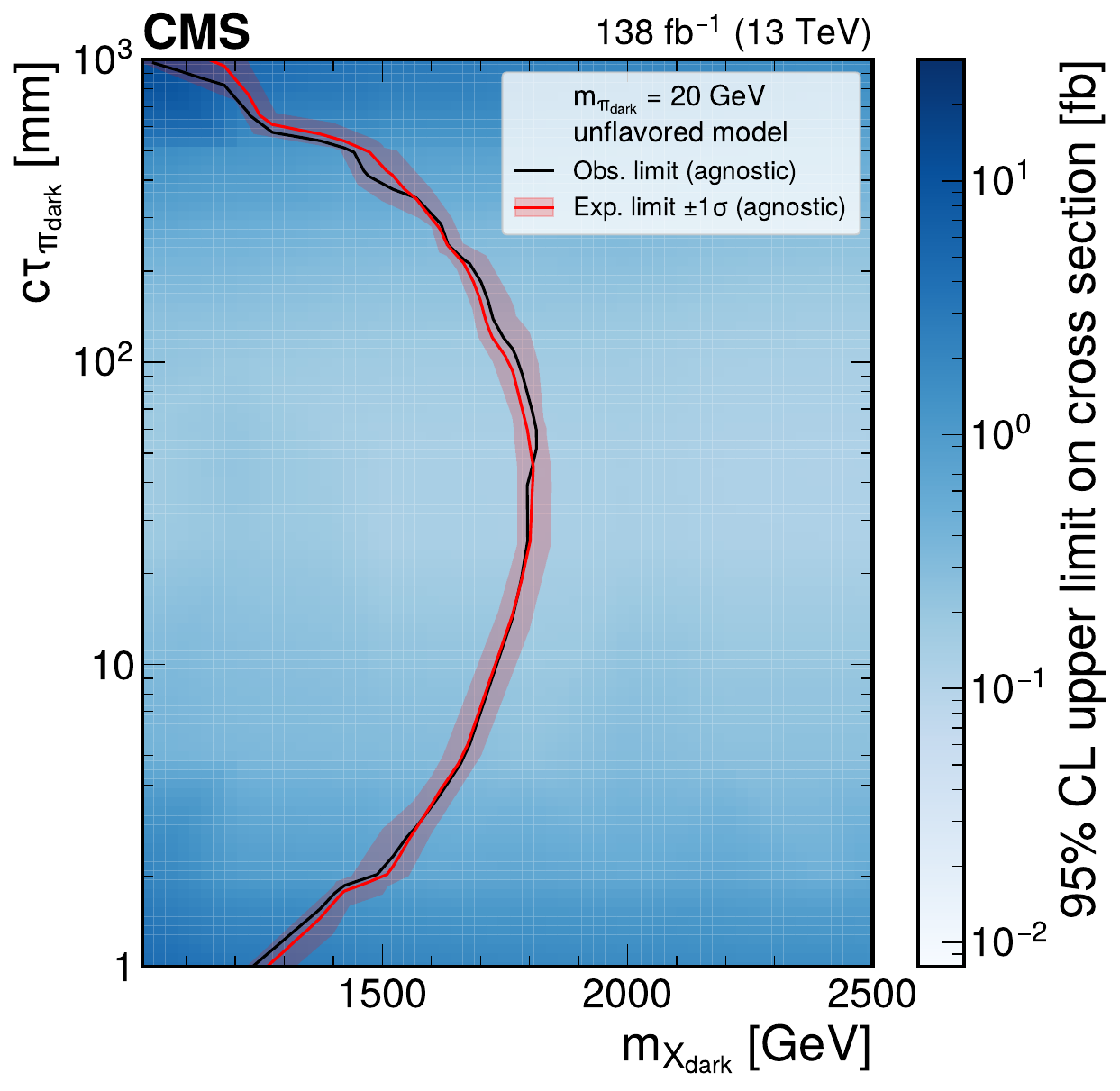}
    \hspace{0.02\textwidth}
    \includegraphics[width=0.48\textwidth]{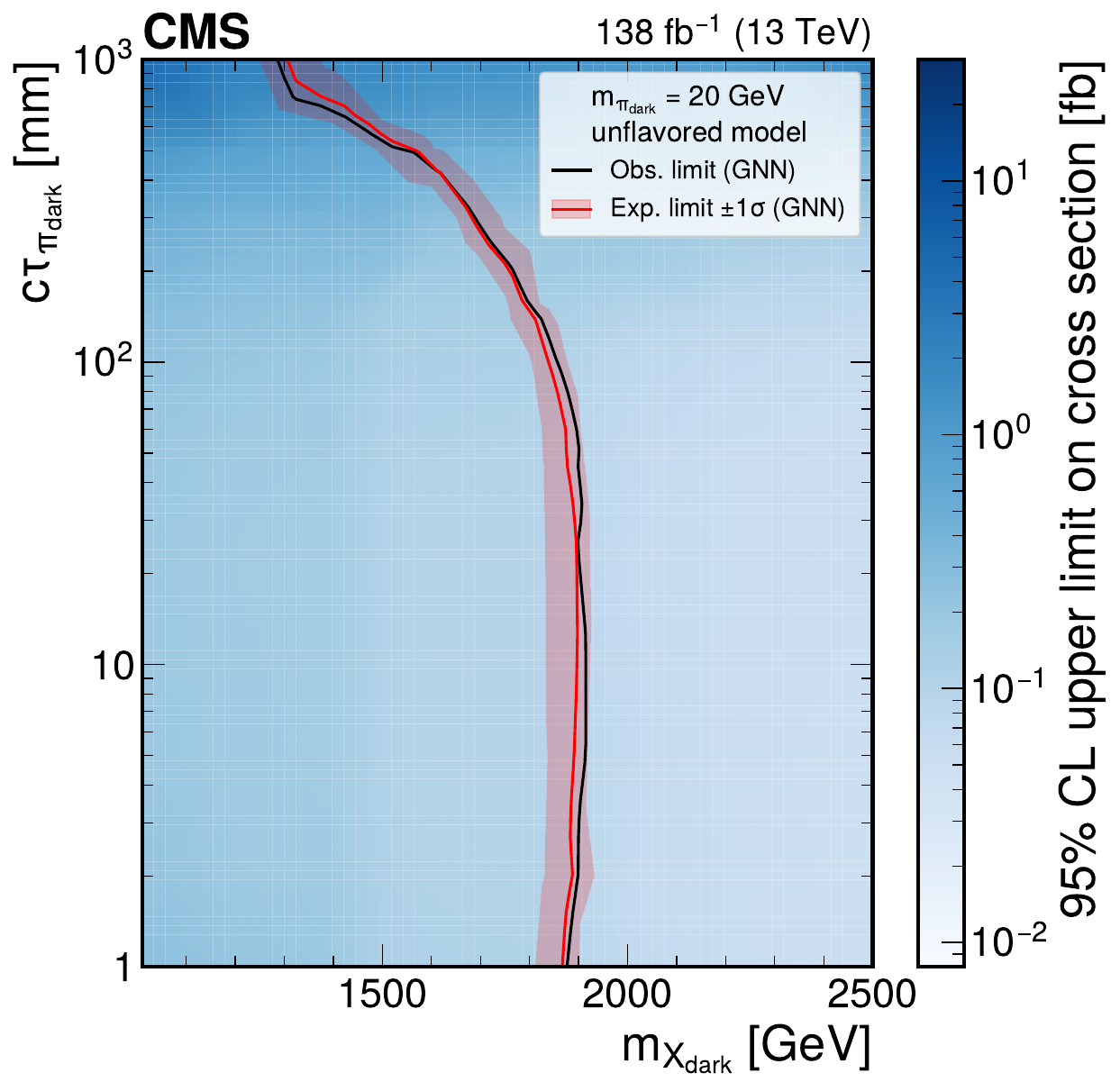} \\
    \includegraphics[width=0.48\textwidth]{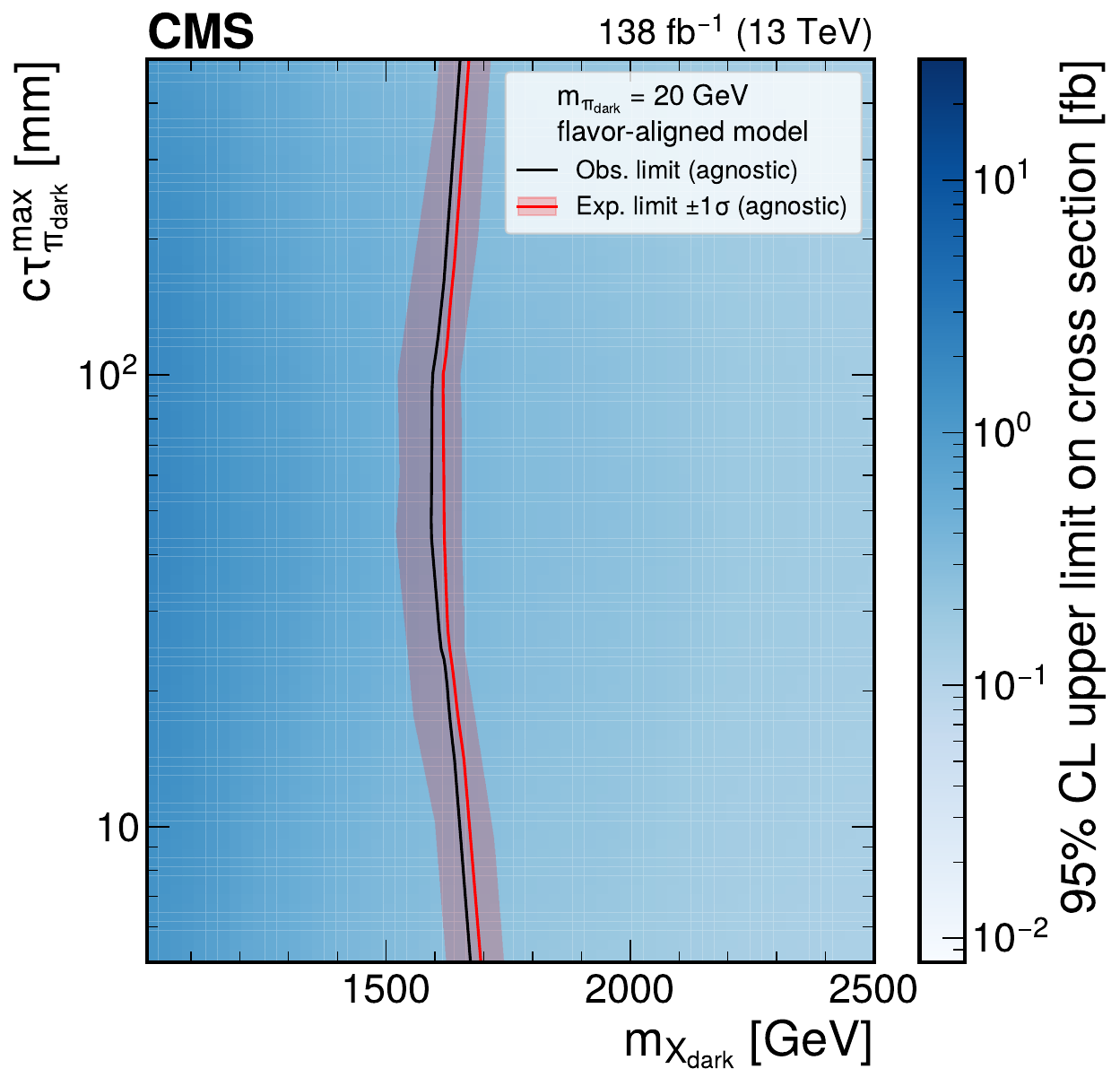}
    \hspace{0.02\textwidth}
    \includegraphics[width=0.48\textwidth]{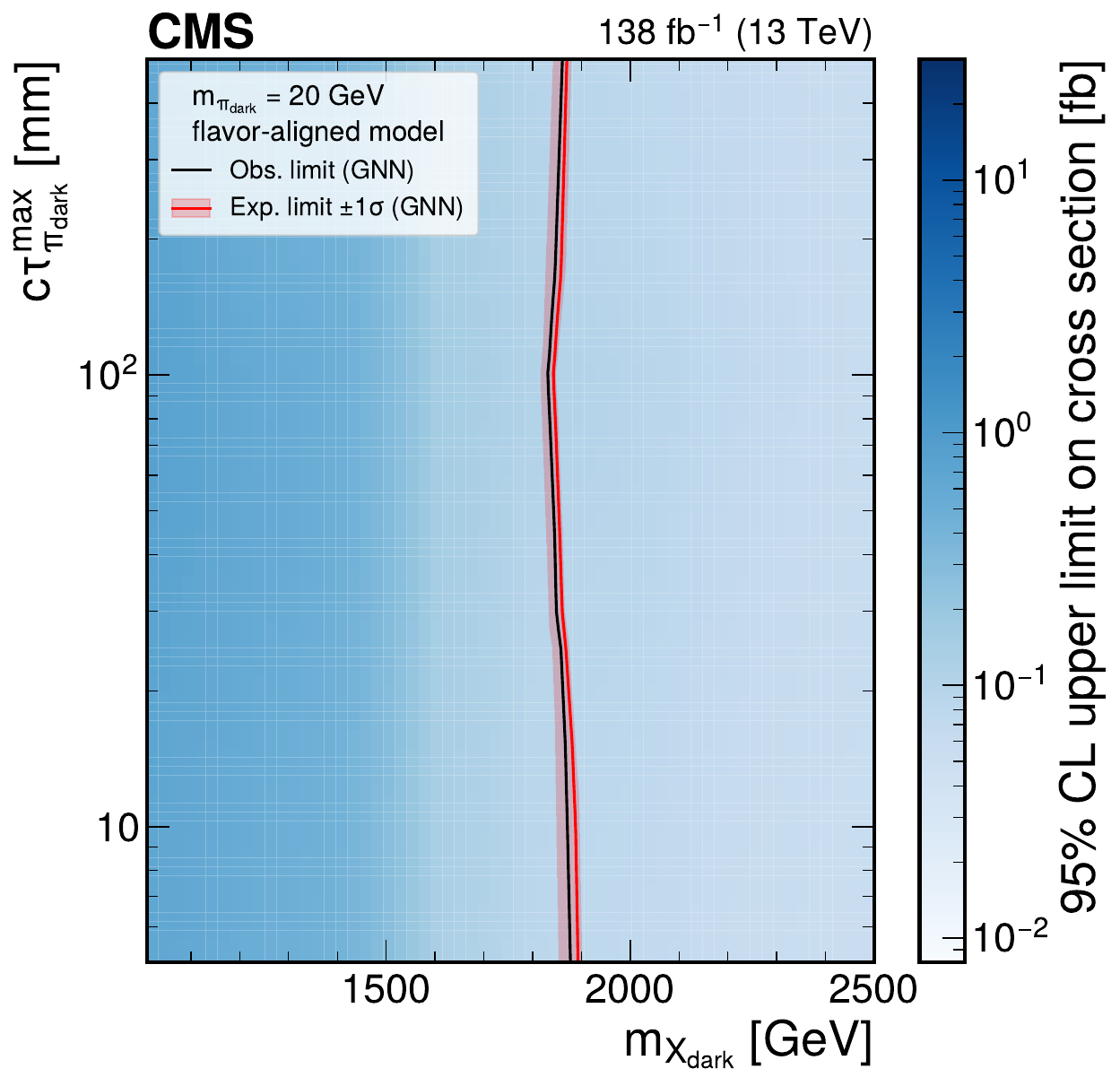}
    \caption{The 95\% \CL upper limits on the production cross section for various signal models in the unflavored scenario (upper plots) and the flavor-aligned scenario (lower plots) with $\mdpi=20\GeV$ using the model-agnostic (GNN) EJ tagging method, on the left (right). The red curve is the expected exclusion limit, with the band representing its 68\% \CL variation. The black curve is the observed limit.} 
    \label{fig:result_mpi20}
\end{figure}

\begin{figure}[tb]
    \centering
    \includegraphics[width=0.48\textwidth]{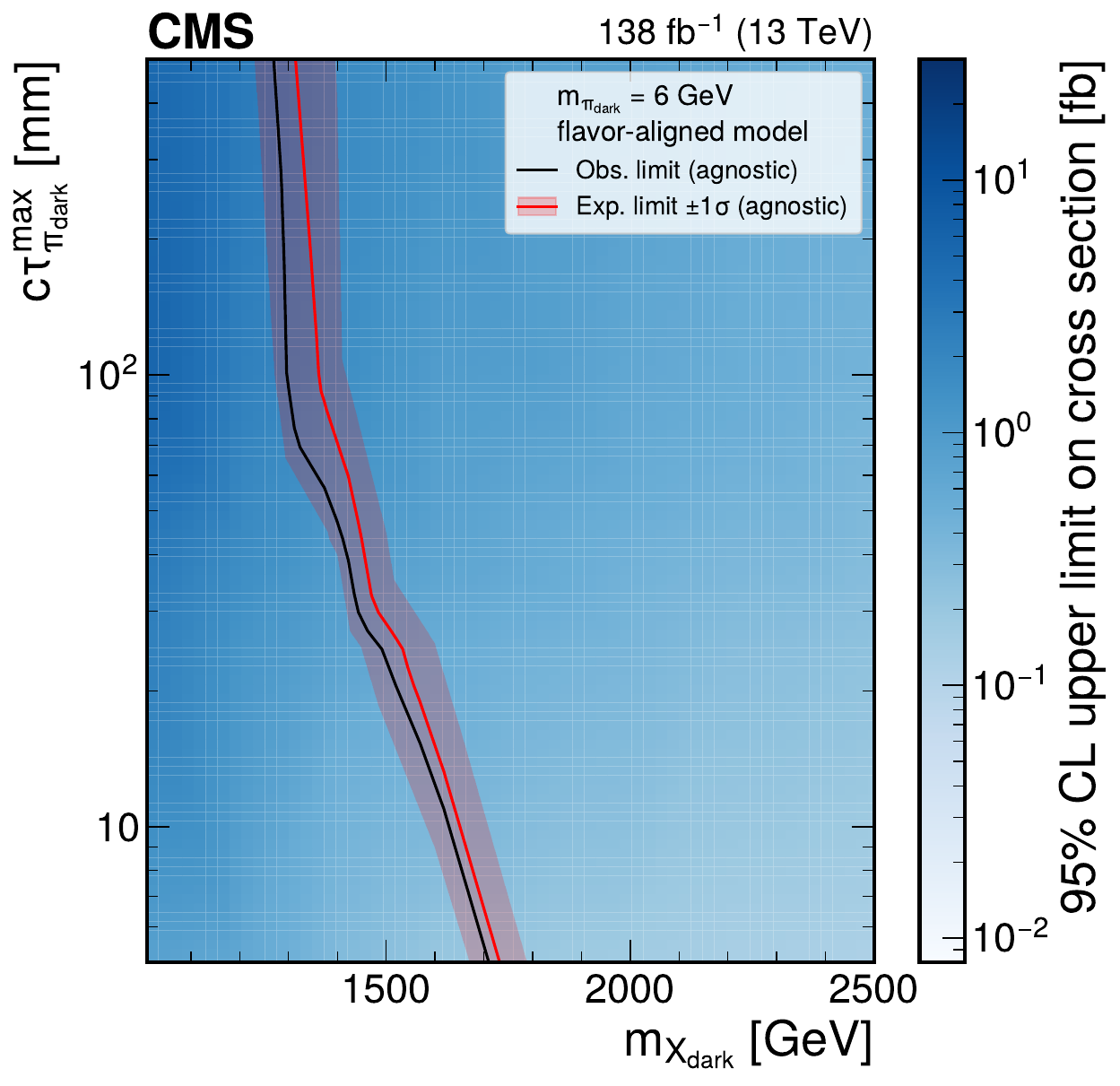}
    \hspace{0.02\textwidth}
    \includegraphics[width=0.48\textwidth]{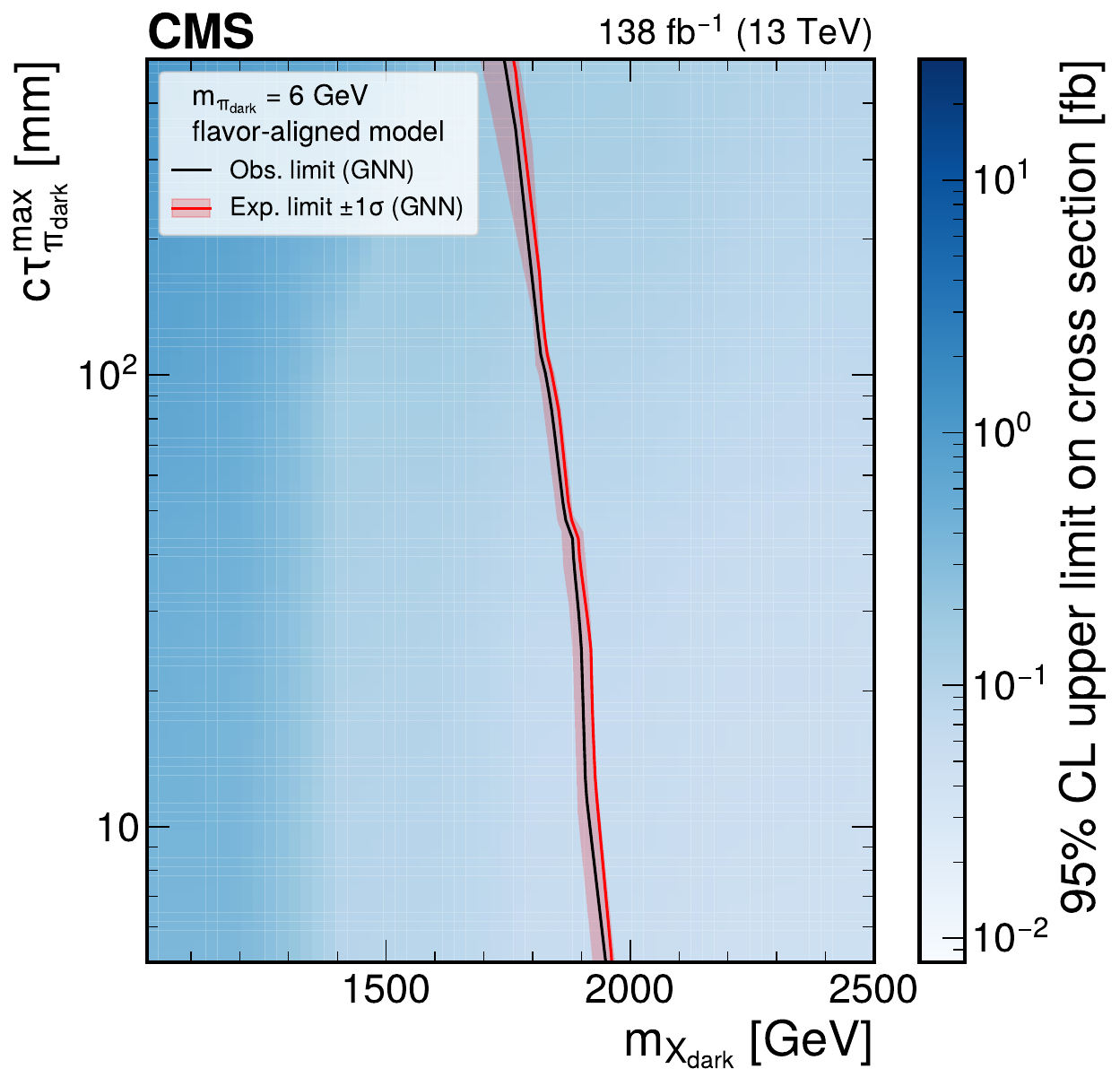}
    \caption{The 95\% \CL upper limits on the production cross section for various signal models in the flavor-aligned scenario with $\mdpi=6\GeV$ using the model-agnostic (GNN) EJ tagging method, on the left (right). The red curve is the expected exclusion limit, with the band representing its 68\% \CL variation. The black curve is the observed limit.}
    \label{fig:result_mpi6}
\end{figure}

After the full event selection, the observed and expected event yields corresponding to each selection set are shown in Table~\ref{tab:eventyield}. No statistically significant deviation from the SM background prediction is observed. The observation is used to set upper limits on the various signal parameter models considered using the \CLs criterion~\cite{Junk:1999kv,Read:2002hq}. The test statistic is defined as the likelihood ratio employed in Higgs boson analyses in ATLAS and CMS, as elaborated in Ref.~\cite{CMS-NOTE-2011-005}.
Upper limits at the 95\% confidence level (\CL) on the production cross section in the 2-dimensional plane defined by the signal parameters are shown in Figs.~\ref{fig:result_mpi10} (\mdpi = 10\GeV) and \ref{fig:result_mpi20} (\mdpi = 20\GeV) for both the unflavored and the flavor-aligned scenarios, and in Fig.~\ref{fig:result_mpi6} ($\mdpi=6\GeV$) for the flavor-aligned scenario.  These excluded cross sections are compared to the theoretical prediction to derive the exclusion regions shown as red and black curves for the expected and observed 95\% \CL limit, respectively.  The red curves also have an associated red band to indicate the 68\% \CL variation on the expected exclusion limit.

\begin{table}[htb]
    \centering
    \topcaption{The estimated number of events from the background prediction based on control samples in data and the observed event yields. Statistical and systematic uncertainties in the estimated background are provided.}
    \renewcommand{\arraystretch}{1.4}
    \label{tab:eventyield}
    \begin{tabular}{cHQ}
        \multicolumn{1}{c}{Selection set} &
        \multicolumn{2}{r@{\symmdelim}}{\TabHeaderA} &
        \multicolumn{2}{@{\hspace{0em}}c@{\symmdelim}}{\TabHeaderB} &
        \multicolumn{2}{@{\hspace{0em}}l}{\TabHeaderC} &
        \multicolumn{1}{c}{\TabHeaderO} \\ [\cmsTabSkip]
    \hline
    u-set 1  &   56&   & $\leftscript{5}{9}$ & $\leftscript{}{}$ & 20&  &  67 \\
    u-set 2  &   20&.0 & $\leftscript{2}{4}$ & $\leftscript{.5}{.3}$ &  7&.0 &  21 \\
    u-set 3  &   22&.9 & $\leftscript{2}{7}$ & $\leftscript{.1}{.3}$ &  4&.9 &  24 \\
    u-set 4  &    7&.9 & $\leftscript{1}{2}$ & $\leftscript{.6}{.0}$ &  2&.2 &  10 \\
    u-set 5  &   11&.3 & $\leftscript{1}{2}$ & $\leftscript{.9}{.7}$ &  2&.0 &  13 \\
    [\cmsTabSkip]
    a-set 1  &    8&.8 & $\leftscript{1}{2}$ & $\leftscript{.0  }{.4}$ &  2&.0 &  16 \\
    a-set 2  &   1&.67 & $\leftscript{0}{0}$ & $\leftscript{.23}{.49}$ & 0&.38 &   3 \\
    a-set 3  &   1&.97 & $\leftscript{0}{0}$ & $\leftscript{.22}{.47}$ & 0&.37 &   2 \\
    a-set 4  &   2&.30 & $\leftscript{0}{0}$ & $\leftscript{.30}{.81}$ & 0&.39 &   3 \\
    a-set 5  &   10&.2 & $\leftscript{1}{2}$ & $\leftscript{.1  }{.3}$ & 3&.4  &  16 \\
    [\cmsTabSkip]
    uGNN set 1  &  15&.6 & $\leftscript{1}{5}$ & $\leftscript{.9  }{.4}$ & 3&.8 &  18 \\
    uGNN set 2  &  0&.73 & $\leftscript{0}{0}$ & $\leftscript{.16}{.44}$ & 0&.27 &  0 \\
    uGNN set 3  &   7&.6 & $\leftscript{1}{3}$ & $\leftscript{.3}{.5}$ & 2&.3 &     9 \\
    [\cmsTabSkip] 
    aGNN set 1  &   45&  & $\leftscript{8}{18}$ & $\leftscript{}{}$      & 16&    &  59\\
    aGNN set 2  &   0&.30 & $\leftscript{0}{ 0}$ & $\leftscript{.07}{.23}$ &  0&.18 &   1\\
    aGNN set 3  &    3&.8 & $\leftscript{0}{ 2}$ & $\leftscript{.7  }{.2}$ &  2&.0  &   5\\
    \end{tabular}
\end{table}

In the unflavored coupling model, the key tagging variable for the model-agnostic method is the mean \trkdxy of the associated tracks in the jet. In the case where the dark pion lifetime is too short, the displaced tracks from the dark pion decay products will be very similar to prompt SM tracks. In the opposite case, where the dark pion lifetime is very large, the dark pions increasingly decay outside the tracker. Thus, this search method is less sensitive to these two cases, giving reduced  performance, as shown in the upper-left plots of Figs.~\ref{fig:result_mpi10} and \ref{fig:result_mpi20}. In contrast, the GNN method performs well even at low dark pion lifetimes, as the GNN leverages information from track-level relationships within EJs to keep the signal acceptance high. However, signal sensitivity is still limited for long dark pion lifetimes, as shown in the upper-right plots of Figs.~\ref{fig:result_mpi10} and \ref{fig:result_mpi20}.

In the flavor-aligned scenario, the differentiating power for the model-agnostic method comes mainly from the multiplicity of associated tracks with large displacements. Unlike \trkdxy, the multiplicity distribution is more uniform for different dark pion lifetimes (as we only consider \ctaumax above the typical \PQb hadron lifetime) leading to less dependence on \ctaumax as seen in the lower-left plots of Figs.~\ref{fig:result_mpi10} and \ref{fig:result_mpi20}. Similarly, the GNN sensitivity is less dependent on changes in lifetime, resulting in a stable signal acceptance across the \ctaumax range. In the flavor-aligned models where $\mdpi=6\GeV$, the dark pions decay predominantly to light SM quarks because of mass constraints, resulting in fewer displaced tracks in the events. This reduces the selection efficiencies for these signal models, leading to the model-agnostic method having less sensitivity compared to models with larger \mdpi, as shown in Fig.~\ref{fig:result_mpi6}.

The ML-based EJ tagging methods yield more stringent limits on the surveyed models. In the unflavored model, \dmed masses up to 1900 and 1950\GeV are excluded at 95\% \CL for \mdpi of 10 and 20\GeV, respectively, closely matching the expected limits.  For the flavor-aligned models, the \dmed mass exclusion at 95\% \CL increases to 1950, 1850, and 1900\GeV for \mdpi of 6, 10, and 20\GeV, respectively, again matching well the expected limits.  Relative to the model-agnostic tagger, the greatest increase in sensitivity is for the unflavored signal models with short lifetimes, where the dark showers generate nearly prompt tracks that are difficult to tag using simple requirements on the track displacement variables.

For the flavor-aligned scenario in the model-agnostic approach, the exclusion range on \mdmed initially decreases as \ctaumax increases. This is due to the decline in signal selection efficiency caused by the loss in track reconstruction efficiency, particularly for more displaced tracks. As \ctaumax continues to increase, the displacement of the dark pion with the next longest lifetime, calculated from Eq.~\eqref{eq:ctau_flavored}, becomes significant for models with $\mdpi=10$ \GeV, resulting in a recovery in signal selection efficiency. However, for models with $\mdpi=6$ \GeV, the decay of dark pions to $\PQb \bar{\PQb}$ is constrained due to mass limitations. Consequently, the fraction of long-lived dark pions is higher compared to models with $\mdpi=10$ \GeV, leading to decreasing signal efficiency even with longer lifetimes.

The GNN-based limits for the flavor-aligned models that utilize the aGNN set 2 selection have a narrow expected band, as shown in the lower right plots of Figs.~\ref{fig:result_mpi10}--\ref{fig:result_mpi20} and the right plot of Fig.~\ref{fig:result_mpi6}.
For that selection set, 0.3 background events are predicted, which implies that the
most frequent observation will be exactly zero events if the background-only hypothesis is correct. Because the likelihood function is constructed from Poisson probabilities, this leads to the background-only test statistic distribution exhibiting discrete, narrow peaks. With a free-floating signal strength, the test statistic distribution for the signal plus background hypothesis shares similar features. As a result, only a small change in the signal strength is required to obtain a \CLs value of 0.05 when going from 95 to 68\% \CL to the expected median of the background expectation. The resulting expected limit bands in this region are narrow and asymmetrical.

\section{Summary}
\label{sec:conclusion}

A search for emerging jet signatures arising from a strongly interacting dark sector produced in proton-proton collisions has been presented, using data corresponding to an integrated luminosity of 138\fbinv at $\sqrt{s}=13\TeV$. The signal model contains a family of dark quarks that couple to the standard model (SM) quarks via a scalar mediator \dmed. Dark pions (\dpi) with a significant lifetime (\ctaudpi) are produced by the hadronization of the dark quarks; these then decay to SM particles at vertices displaced from the proton-proton interaction point. As the scalar mediator is assumed to be produced in pairs, and each decays to an SM quark and a dark quark, the signature of this process is two SM jets plus two jets of particles with constituents emerging from displaced vertices.

Both unflavored and flavor-aligned couplings between the SM quarks and the dark quarks are examined in the search.
Events are selected using either a traditional cut-based approach or a graph neural network to identify emerging jets, in combination with other event-level selection criteria. The overall selection requirements are optimized for each coupling scenario and for different combinations of the mediator particle mass, dark pion mass, and dark pion lifetime. 
No excess of events beyond the SM expectations is found, and the observed 95\% confidence level exclusion limits agree with the expected limits.  For the unflavored model, dark mediator masses $\mdmed<1950\GeV$ are excluded for $\ctaudpi\approx100\mm$ and $\mdpi=10\GeV$, while the flavor-aligned model result excludes $\mdmed<1850\GeV$ at $\ctaumax\approx500\mm$ for $\mdpi=10\GeV$. This result surpasses the previous search for emerging jets in the unflavored scenario, increasing the experimental limit of the dark mediator particle by ${\approx}500\GeV$ to set the most stringent limits to date, and provides the first direct exclusion of the flavor-aligned scenario. 

\begin{acknowledgments}
We congratulate our colleagues in the CERN accelerator departments for the excellent performance of the LHC and thank the technical and administrative staffs at CERN and at other CMS institutes for their contributions to the success of the CMS effort. In addition, we gratefully acknowledge the computing centers and personnel of the Worldwide LHC Computing Grid and other centers for delivering so effectively the computing infrastructure essential to our analyses. Finally, we acknowledge the enduring support for the construction and operation of the LHC, the CMS detector, and the supporting computing infrastructure provided by the following funding agencies: SC (Armenia), BMBWF and FWF (Austria); FNRS and FWO (Belgium); CNPq, CAPES, FAPERJ, FAPERGS, and FAPESP (Brazil); MES and BNSF (Bulgaria); CERN; CAS, MoST, and NSFC (China); MINCIENCIAS (Colombia); MSES and CSF (Croatia); RIF (Cyprus); SENESCYT (Ecuador); ERC PRG, RVTT3 and MoER TK202 (Estonia); Academy of Finland, MEC, and HIP (Finland); CEA and CNRS/IN2P3 (France); SRNSF (Georgia); BMBF, DFG, and HGF (Germany); GSRI (Greece); NKFIH (Hungary); DAE and DST (India); IPM (Iran); SFI (Ireland); INFN (Italy); MSIP and NRF (Republic of Korea); MES (Latvia); LMTLT (Lithuania); MOE and UM (Malaysia); BUAP, CINVESTAV, CONACYT, LNS, SEP, and UASLP-FAI (Mexico); MOS (Montenegro); MBIE (New Zealand); PAEC (Pakistan); MES and NSC (Poland); FCT (Portugal); MESTD (Serbia); MCIN/AEI and PCTI (Spain); MOSTR (Sri Lanka); Swiss Funding Agencies (Switzerland); MST (Taipei); MHESI and NSTDA (Thailand); TUBITAK and TENMAK (Turkey); NASU (Ukraine); STFC (United Kingdom); DOE and NSF (USA).

\hyphenation{Rachada-pisek} Individuals have received support from the Marie-Curie program and the European Research Council and Horizon 2020 Grant, contract Nos.\ 675440, 724704, 752730, 758316, 765710, 824093, 101115353, and COST Action CA16108 (European Union); the Leventis Foundation; the Alfred P.\ Sloan Foundation; the Alexander von Humboldt Foundation; the Science Committee, project no. 22rl-037 (Armenia); the Belgian Federal Science Policy Office; the Fonds pour la Formation \`a la Recherche dans l'Industrie et dans l'Agriculture (FRIA-Belgium); the Agentschap voor Innovatie door Wetenschap en Technologie (IWT-Belgium); the F.R.S.-FNRS and FWO (Belgium) under the ``Excellence of Science -- EOS" -- be.h project n.\ 30820817; the Beijing Municipal Science \& Technology Commission, No. Z191100007219010 and Fundamental Research Funds for the Central Universities (China); the Ministry of Education, Youth and Sports (MEYS) of the Czech Republic; the Shota Rustaveli National Science Foundation, grant FR-22-985 (Georgia); the Deutsche Forschungsgemeinschaft (DFG), under Germany's Excellence Strategy -- EXC 2121 ``Quantum Universe" -- 390833306, and under project number 400140256 - GRK2497; the Hellenic Foundation for Research and Innovation (HFRI), Project Number 2288 (Greece); the Hungarian Academy of Sciences, the New National Excellence Program - \'UNKP, the NKFIH research grants K 124845, K 124850, K 128713, K 128786, K 129058, K 131991, K 133046, K 138136, K 143460, K 143477, 2020-2.2.1-ED-2021-00181, and TKP2021-NKTA-64 (Hungary); the Council of Science and Industrial Research, India; ICSC -- National Research Center for High Performance Computing, Big Data and Quantum Computing, funded by the NextGenerationEU program (Italy); the Latvian Council of Science; the Ministry of Education and Science, project no. 2022/WK/14, and the National Science Center, contracts Opus 2021/41/B/ST2/01369 and 2021/43/B/ST2/01552 (Poland); the Funda\c{c}\~ao para a Ci\^encia e a Tecnologia, grant CEECIND/01334/2018 (Portugal); the National Priorities Research Program by Qatar National Research Fund; MCIN/AEI/10.13039/501100011033, ERDF ``a way of making Europe", and the Programa Estatal de Fomento de la Investigaci{\'o}n Cient{\'i}fica y T{\'e}cnica de Excelencia Mar\'{\i}a de Maeztu, grant MDM-2017-0765 and Programa Severo Ochoa del Principado de Asturias (Spain); the Chulalongkorn Academic into Its 2nd Century Project Advancement Project, and the National Science, Research and Innovation Fund via the Program Management Unit for Human Resources \& Institutional Development, Research and Innovation, grant B37G660013 (Thailand); the Kavli Foundation; the Nvidia Corporation; the SuperMicro Corporation; the Welch Foundation, contract C-1845; and the Weston Havens Foundation (USA).
\end{acknowledgments}

\bibliography{auto_generated}
\cleardoublepage \appendix\section{The CMS Collaboration \label{app:collab}}\begin{sloppypar}\hyphenpenalty=5000\widowpenalty=500\clubpenalty=5000
\cmsinstitute{Yerevan Physics Institute, Yerevan, Armenia}
{\tolerance=6000
A.~Hayrapetyan, A.~Tumasyan\cmsAuthorMark{1}\cmsorcid{0009-0000-0684-6742}
\par}
\cmsinstitute{Institut f\"{u}r Hochenergiephysik, Vienna, Austria}
{\tolerance=6000
W.~Adam\cmsorcid{0000-0001-9099-4341}, J.W.~Andrejkovic, T.~Bergauer\cmsorcid{0000-0002-5786-0293}, S.~Chatterjee\cmsorcid{0000-0003-2660-0349}, K.~Damanakis\cmsorcid{0000-0001-5389-2872}, M.~Dragicevic\cmsorcid{0000-0003-1967-6783}, P.S.~Hussain\cmsorcid{0000-0002-4825-5278}, M.~Jeitler\cmsAuthorMark{2}\cmsorcid{0000-0002-5141-9560}, N.~Krammer\cmsorcid{0000-0002-0548-0985}, A.~Li\cmsorcid{0000-0002-4547-116X}, D.~Liko\cmsorcid{0000-0002-3380-473X}, I.~Mikulec\cmsorcid{0000-0003-0385-2746}, J.~Schieck\cmsAuthorMark{2}\cmsorcid{0000-0002-1058-8093}, R.~Sch\"{o}fbeck\cmsorcid{0000-0002-2332-8784}, D.~Schwarz\cmsorcid{0000-0002-3821-7331}, M.~Sonawane\cmsorcid{0000-0003-0510-7010}, S.~Templ\cmsorcid{0000-0003-3137-5692}, W.~Waltenberger\cmsorcid{0000-0002-6215-7228}, C.-E.~Wulz\cmsAuthorMark{2}\cmsorcid{0000-0001-9226-5812}
% ...
\par}
\vskip\cmsinstskip
\dag:~Deceased\\
$^{1}$Also at Yerevan State University, Yerevan, Armenia\\
$^{2}$Also at TU Wien, Vienna, Austria\\
% ...
\end{sloppypar}
%%% END EDITABLE REGION %%%
% skeleton_end
\end{document}